\documentclass{ws-m3as}
\usepackage{}
\usepackage{amsfonts}
\usepackage{mathbbold}
\usepackage{mathrsfs}

\begin{document}

\markboth{S.Q. Wang, W.H. Deng, Y.J. Wu, J.Y. Yuan}{CLDGM for solving time-dependent convection-dominated NS Eq.}

%%%%%%%%%%%%%%%%%%% Publisher's Area please ignore %%%%%%%%%%%%%%%%%%%%%%%
%
\catchline{}{}{}{}{}

%%%%%%%%%%%%%%%%%%%%%%%%%%%%%%%%%%%%%%%%%%%%%%%%%%%%%%%%%%%%%%%%%%%%%%%%%%

\title{Characteristic local discontinuous Galerkin methods for solving time-dependent convection-dominated
 Navier-Stokes equations\footnote{This work was partially supported by the National Basic Research (973) Program
 of China under Grant 2011CB706903, the National Natural Science Foundation of China under Grant 11271173, the
 Fundamental Research Funds for Central Universities under Grant lzujbky-2012-14, and the CAPES and CnPq in Brazil.} }

\author{Shuqin Wang%\footnote{}%
}

\address{School of Mathematics and Statistics, Lanzhou University, Lanzhou 730000,
P.R. China\\\
Department of Mathematics, Federal University of Parana, Centro Politecnico, Curitiba, CEP: 81531-990, PR, Brazil %\footnote{}%
\\
wsqlzu@gmail.com}

\author{Weihua Deng\footnote{Corresponding author.}}
\address{School of Mathematics and Statistics, Lanzhou University, Lanzhou 730000, P.R. China\\
dengwh@lzu.edu.cn}

\author{Yujiang Wu}
\address{School of Mathematics and Statistics, Lanzhou University, Lanzhou 730000, P.R. China\\
myjaw@lzu.edu.cn}

\author{Jinyun Yuan}
\address{Department of Mathematics, Federal University of Parana, Centro Politecnico, Curitiba, CEP: 81531-990, PR, Brazil\\
yuanjy@gmail.com}

\maketitle

%\begin{history}
%\received{(Day Month Year)}
%\revised{(Day Month Year)}
%\accepted{(Day Month Year)}
%\comby{(xxxxxxxxxx)}
%\end{history}

\begin{abstract}
Combining the characteristic method and the local discontinuous
Galerkin method with carefully constructing numerical fluxes, we
design the variational formulations for the time-dependent
convection-dominated Navier-Stokes equations in $\mathbb{R}^2$. The
proposed symmetric variational formulation is strictly proved to be
unconditionally stable; and the scheme has the striking benefit that the conditional number of the matrix of the corresponding matrix equation does not increase with the refining of the meshes. The presented scheme works well for a wide
range of Reynolds numbers, e.g., the scheme still has good error
convergence when $Re=0.5 e+005$ or $1.0 e+ 008$. Extensive
numerical experiments are performed to show the optimal convergence
orders and the contours of the solutions of the equation with given
initial and boundary conditions.

%A Characteristic Local Discontinuous Galerkin
%Method is proposed for solving time dependent convection-dominated Navier-Stokes equations in $\mathbb{R}^2$.
%The numerical simulations illustrate that the proposed method is efficient with good error convergence when $Re=0.5\times 10^5$ or $1 \times 10^8$.
%Unconditional stability is obtained by the symmetric variational formula. Finally some numerical examples confirm the theoretical predictions.
\end{abstract}

\keywords{Time-dependent Navier-Stokes equations; local discontinuous Galerkin method; symmetric variational formulation.}

\ccode{AMS Subject Classification: 35Q30, 65M60}

\section{Introduction}
Based on the assumption that the fluid, at the scale of interest, is a continuum, and the conservation of momentum (often alongside mass and energy conservation),
the equation to describe the motion of fluid substances can be derived, which is named after the French engineer and physicist Claude-Louis Navier and
the Ireland mathematician and physicist George Gabriel Stokes to memory their fundamental contributions.
Nowadays, it is still the central equation to fluid mechanics.
Let $\Omega$ be a bounded polygonal domain in $\mathbb{R}^2$ with Lipschitz-continuous boundary $\partial\Omega$.
The time-dependent Navier-Stokes equation for an incompressible viscous fluid confined in $\Omega$ is \cite{r16}:

\begin{equation} \label{1.1}
\left\{ \begin{array}
 {l@{\quad} l}
 \bm u_t+(\bm u\cdot\nabla)\bm u-\nu\Delta\bm u+\nabla p=\bm f ,& \ \ (\bm x,t)\in\Omega\times[0,T],\\
 %\cr\noalign{\vskip  0.01 mm}
  \nabla\cdot\bm u=0, & \ \  (\bm x,t)\in\Omega\times[0,T],\\
  \bm u(\bm x,0)=\bm u_0(\bm x), & \ \ \ \bm x\in\Omega,\\
  \bm u(\bm x,t)=0, & \ \  (\bm x,t)\in\partial\Omega\times[0,T].
 \end{array}
 \right.
\end{equation}

It is well known that if the body force function $\bm f\in
L^2(0,T;H^{-1}(\Omega)^2)$ and  the initial value $\bm u_0\in
H(div,\Omega)$, then this problem has a unique solution $\bm u\in
L^2(0,T;H^1_0(\Omega)^2)\cap L^{\infty}(0,T;L_0^2(\Omega))$, $p\in
W^{-1,\infty}(0,T;L_0^2(\Omega))$, and $\bm u_t\in L^2(0,T;\bm
V^{\prime})$, where $\bm V$ is defined below (\ref{2.4}) \cite{r16}.
The constant $\nu$ is the fluid viscosity coefficients. Since $p$ is
uniquely defined up to an additive constant, we also assume that
$\int_{\Omega}p=0$. The $ (\bm u\cdot\nabla)\bm u$ is a nonlinear
convective term. If the components of $\bm u$ are $u_1$ and $u_2$,
this term is defined as
$$(\bm u\cdot\nabla)\bm u=u_1\frac{\partial\bm u}{\partial x_1}+u_2\frac{\partial\bm u}{\partial x_2}\ .$$

The idea of the characteristic methods dates back to the works of
Arbogast and Wheeler \cite{r1}, and Boukir et al \cite{r2}. Boukir
et al further extend the idea to two and three dimensional nonlinear
coupled system, and the detailed numerical analysis for the
incompressible Navier-Stokes equation is performed \cite{r3}. The
characteristic method tackles the time derivative term and the
nonlinear convective term together; we use it to solve the
considered equation with first order accuracy in time. It seems that
the advantages of the characteristic methods can be listed as: 1)
efficient in solving the advection-dominated diffusion problems; 2)
easily obtaining the existence and uniqueness of the solutions of
the discretized system; 3) making the nonlinear equations linear,
conveniently tackling the nonlinear obstacles; 4) easily performing
the numerical stability analysis. Another idea to treat the
nonlinear term is to use the technique of operator splitting
\cite{r12}.

Because of the inherent performances of the Navier-Stokes or Stokes
equations in characterizing the turbulence (most flows occurring in
nature are turbulent) in fluids or gases, from the finite element
methods to discontinuous Galerkin methods a lot of research works on
these topics have been done
\cite{r4,r5,r6,r7,r8,r9,r10,r14,r17,r24}; and important progresses
have been made. However, it seems that there are less works on the
discontinuous Galerkin method to solve the time-dependent
incompressible Navier-Stokes equation, and much less works on the
local discontinuous Galerkin method. Splitting the nonlinearity and
incompressibility, and using discontinuous or continuous finite
element methods in space, Girautl et al solve the time-dependent
incompressible Navier-Stokes equation \cite{r12}. In this paper, we
use the local discontinuous Galerkin methods to discretize the space
derivative of the considered equation. It seems that the following
advantages can be obtained: 1) by introducing the local auxiliary
variable, the order of the diffusion term can be reduced and adding
the penalty term makes the symmetric formulation possible which is
valuable for stability analysis and numerical computation; 2) the
introduced auxiliary variable $\bar{\bm\sigma}=\sqrt{\nu}\nabla\bm
u$ lessens the challenges caused by the big Reynold number since
$\sqrt{\nu}$ is not as small as $\nu$. Of course, the general
advantages of discontinuous Galerkin methods still exist, e.g.,
suitable for complex geometries, easy to get high order accuracy and
to perform $hp$-adaptivity, and the semi-discrete scheme is
explicit, etc.

The outline of this paper is as follows. The computational schemes
are presented and discussed in Sec. 2. We prove the existence,
uniqueness, and numerical stability in Sec. 3. We perform the
numerical experiments and show some numerical simulations to verify
the theoretical results and illustrate the powerfulness of the given
schemes in Sec. 4. Some concluding remarks are given in Sec. 5; and
in the Appendix, we present the other two different variational
formulations for the time-dependent Navier-Stokes equations.

%In this work, a Discontinuous Galerkin approximation scheme is described and shown in Section 2. The existence and stability properties are prescribed in Section 3. A suboptimal velocity error estimate and pressure error estimate in $L^2(\Omega)$ space are obtained in Section 4. In Section 5, we will give numerical experiment and show the computational details. At last, in Appendix we will give two other variational formulas and their properties.
%

\section{Derivation of the numerical scheme}

We first introduce the notations, and then focus on deriving the full discrete numerical schemes of the time-dependent Navier-Stokes equations.

\subsection{Preliminaries}

For the mathematical setting of the Navier-Stokes problems, we
describe some Sobolev spaces. The $L^2(\Omega)$ is the classical
space of square integrable functions with the inner product
$(f,g)=\int_{\Omega}fg\ dx$; and $L_0^2(\Omega)$ is the subspace of
functions of $L^2(\Omega)$ with zero mean value, that is,
     $$L_0^2(\Omega)=\left\{v\in L^2(\Omega):\int_{\Omega}v=0\right\};$$
and %$H^1(\Omega)$ denotes the classical Sobolev space as
       $$H^1(\Omega)=\left\{v\in L^2(\Omega):\nabla v\in L^2(\Omega)\right\}.$$
It is well known that $C_0^{\infty}(\Omega)$ is the space of
infinitely differentiable functions with compact support; and
$H^1_0(\Omega)$ is the closure of $C_0^{\infty}(\Omega)$ in
$H^1(\Omega)$. The $H^{-1}(\Omega)$ is the dual space of
$H^1_0(\Omega)$. Denote $\bm X$ as the space of functions of
$(H^1_0(\Omega))^2$ with zero divergence, i.e.,
$$\bm X=\left\{ \bm v\in(H_0^1(\Omega))^2: \nabla\cdot\bm v=0 \right\},$$
and $\bm X^{\prime}$ as its dual space. The fundamental work spaces for solving the Navier-Stokes equations are $\bm X$ and $M$:
$$M:=L_0^2(\Omega)=\left\{v\in L^2(\Omega):\,\int_{\Omega}v dx=0\right\}. $$
The inner product and norm of vector functions $\bm v=(v_i)_{1\leq i\leq d}$ are defined as:
$$(\bm u,\bm v)_{\Omega}=\int_{\Omega}\bm u\cdot\bm v\ ,\ \ \parallel\bm v\parallel_{L^2(\Omega)}=\Big(\sum_{i=1}^{d}\parallel v_i\parallel_{L^2(\Omega)}^2\Big)^{1/2}.$$
%The following are about the vector notations.
The gradient of a
vector function $\bm v :\mathbb{R}^d \rightarrow \mathbb{R}^d$ is a
matrix; and the divergence of a matrix function $\bm
A:\mathbb{R}^d\rightarrow\mathbb{R}^{d\times d}$ is a vector:
$$\nabla\bm v=\Big(\frac{\partial v_i}{\partial x_j}\Big)_{1\leq i,j\leq d}\ ,\ \ \ \ \nabla\cdot\bm A=\Big(\sum_{j=1}^d\frac{\partial A_{ij}}{\partial x_j}\Big)_{1\leq i,j\leq d}.$$
Consequently, for a vector function $\bm v=(v_i)_{1\leq i\leq d}$, we have
$$\Delta\bm v=\nabla\cdot\nabla\bm v=(\Delta v_i)_{1\leq i\leq d}.$$
The $L^2$ inner product of two matrix functions $\bm A,\bm B$ is defined by
$$(\bm A,\bm B)_{\Omega}=\int_{\Omega}\bm A:\bm B=\int_{\Omega}\sum_{1\leq i,j\leq d}\bm A_{ij}\bm B_{ij},$$
equipped with the norm
$$\parallel\bm A\parallel=(\bm A,\bm A)_{\Omega}^{1/2}=\Big(\int_{\Omega}\bm A:\bm A\Big)^{1/2}=\Big(\int_{\Omega}\sum_{1\leq i,j\leq d}A_{ij}^2\Big)^{1/2}.$$
Obviously, it is a norm; we just prove that it possesses the third
property of a norm as follows:
\begin{align*}
\begin{split}
\parallel\bm A+\bm B\parallel^2&=\int_{\Omega}(\bm A+\bm B):(\bm A+\bm B) \\ %,\ \forall \bm A\ ,\bm B\\
&=\int_{\Omega}\sum_{1\leq i,j\leq d}(A_{ij}+B_{ij})^2\\
&=\int_{\Omega}\sum_{1\leq i,j\leq d}(A_{ij}^2+2A_{ij}B_{ij}+B_{ij}^2)\\
&=\parallel\bm A\parallel^2+2(\bm A,\bm B)_{\Omega}+\parallel \bm B\parallel^2\\
&\leq\parallel\bm A\parallel^2+2\parallel\bm A\parallel\ \parallel\bm B\parallel+\parallel\bm B\parallel^2\\
&\leq(\parallel\bm A\parallel+\parallel\bm B\parallel)^2,
\end{split}
\end{align*}

since%

\begin{align*}
\begin{split}
(\bm A,\bm B)_{\Omega}&=\int_{\Omega}\sum_{1\leq i,j\leq d}A_{ij}B_{ij}\\
&\leq\int_{\Omega}\big(\sum_{1\leq i,j\leq d}A_{ij}^2\big)^{1/2}\big(\sum_{1\leq i,j\leq d}B_{ij}^2\big)^{1/2}\\
&\leq\big(\int_{\Omega}\sum_{1\leq i,j\leq d}A_{ij}^2\big)^{1/2}\big(\int_{\Omega}\sum_{1\leq i,j\leq d}B_{ij}^2\big)^{1/2}\\
&=\parallel\bm A\parallel\ \parallel\bm B\parallel.
\end{split}
\end{align*}

The Broken Sobolev spaces are the natural spaces to work with the DG methods. These spaces depend strongly on the partition of the domain. Let $\Omega$ be a polygonal domain subdivided into elements $E$. Here $E$ is a triangle or a quadrilateral in 2D. We assume that the intersection of two elements is either empty, or an edge (2D). The mesh is called a regular mesh if
$$\forall E\in\mathscr{E}_h,\ \ \ \ \ \ \ \frac{h_E}{\rho_E}\leq C,$$
where $\mathscr{E}_h$ is the subdivision of $\Omega$, $C$ is a constant, $h_E$ is the diameter of the element $E$, and $\rho_E$ is the diameter of the inscribed circle in element $E$.

We introduce the Broken Sobolev space for any real functions,
$$H^s(\mathscr{E}_h)^2=\left\{\bm v\in L^2(\Omega)^2:\,\forall E\in\mathscr{E}_h,\bm v|_E\in H^s(E)^2\right\},$$
equipped with the Broken Sobolev norm:
$$\parallel \bm v\parallel_{H^s(\mathscr{E}_h)}=\big(\sum_{E\in\mathscr{E}_h}\sum_{i=1}^2\parallel v_i\parallel^2_{H^s(E)}\big)^{1/2}.$$

\textbf{Jumps and averages}:
\vskip 0.2cm

We denote by $\mathscr{E}_h^B$ the set of edges of the subdivision $\mathscr{E}_h$. Let  $\mathscr{E}_h^i$ denote the set of interior edges; and $\mathscr{E}_h^b=\mathscr{E}_h^B\ \backslash  \mathscr{E}_h^i$ the set of edges on $\partial\Omega$. With each edge $e$, we have a unit normal vector $n_e$. If $e$ is on the boundary $\partial\Omega$, then $n_e$ is taken to be the unit outward vector normal to $\partial\Omega$.

If $\bm v$ belongs to $H^1(
\mathscr{E}_h)^2$, the trace of $\bm v$ along any side of one element $E$ is well defined. If two elements $E_1^e$ and $E_2^e$ are neighbors and share one common side $e$, there are two traces of $\bm v$ belong to $e$. Now we define an average and
a jump for $\bm v$. We assume that the normal vector $n_e$ is oriented from $E_1^e$ to $E_2^e$, and
$$\{\{\bm v\}\}=\frac{1}{2}(\bm v|_{E_1^e}+\bm v|_{E_2^e}),\ \ \ \lbrack\bm v\rbrack=(\bm v|_{E_1^e}-\bm v|_{E_2^e}), \ \forall e\in\partial E_1^e\bigcap \partial E_2^e.$$
If $e$ is on $\partial\Omega$, we have the definition:
$$\{\{\bm v\}\}=\lbrack\bm v\rbrack=\bm v|_{E},\ \ \ \ \forall e\in\partial E\bigcap\partial\Omega.$$

\subsection{Scheme}

By introducing an auxiliary variable $\bar{\bm\sigma}=\sqrt{\nu}\nabla\bm u$, we rewrite (\ref{1.1}) as a mixed form:
\begin{equation} \label{2.3p}
\left\{ \begin{array}
 {l@{\quad} l}
 \bm u_t+(\bm u\cdot\nabla)\bm u-\sqrt{\nu}\nabla\cdot\bar{\bm\sigma}+\nabla p=\bm f, &
 \ \ \ (\bm x,t) \in\Omega\times[0,T],\\
 %\cr\noalign{\vskip  0.01 mm}
  \bar{\bm\sigma}=\sqrt{\nu}\nabla\bm u, &\ \ \ (\bm x,t) \in\Omega\times[0,T],\\
  \nabla\cdot\bm u=0,&\ \ \ (\bm x,t) \in\Omega\times[0,T],\\
  \bm u(\bm x,0)=\bm u_0(\bm x),&\ \ \ \, \bm x\in\Omega,\\
  \bm u(\bm x,t)=0,&\ \ \ (\bm x,t) \in\partial\Omega\times[0,T],
 \end{array}
 \right.
\end{equation}
where $\nu=1/Re$ is the viscosity coefficient. Obviously, if $\sqrt{\nu}$ is small enough we have  $\sqrt{\nu}>\nu$.

Before presenting the variational form, let us clarify the notation: $\bm v\cdot \bar{\bm\sigma}\cdot\bm n:=\sum_{i,j=1}^2 v_i\bar{\sigma}_{ij}n_j:=\bar{\bm\sigma}:(\bm v\otimes\bm n) $. Multiplying the first, the second, and the
 third equation of (\ref{2.3p}), by the smooth test functions $\bm v, \bar{\bm \tau}, q$, respectively, and integrating by parts over an arbitrary subset $E\in\mathscr{E}_h$, we get the following weak variational formulation:
\begin{equation} \label{2.4}
\left\{ \begin{array}
 {l@{\quad} l}
 \int_E(\bm u_t+(\bm u\cdot\nabla)\bm u)\cdot\bm v+\int_E\sqrt{\nu}\bar{\bm\sigma}:\nabla\bm v-\int_{\partial E}\sqrt{\nu}\bar{\bm\sigma}\cdot\bm v\cdot\bm n_E\\
-\int_E p\nabla\cdot\bm v+\int_{\partial E}p\bm v\cdot \bm n_E=\int_E\bm f\cdot\bm v,\ \ \ \ \ \ \forall \bm v\in\bm V ,\\
\\
\int_E\bar{\bm\sigma}:\bar{\bm\tau}-\int_E\sqrt{\nu}\nabla\bm u:\bar{\bm\tau}=0,\ \ \ \ \ \ \ \ \ \ \ \ \ \ \ \ \forall \bar{\bm\tau}\in\bm V^2,\\
\\
\int_E\nabla\cdot\bm u q=0,\ \ \ \ \ \ \ \ \ \ \ \ \ \ \ \ \ \ \ \ \ \ \ \ \ \ \ \ \ \ \ \ \ \ \ \forall q\in M,
 \end{array}
 \right.
\end{equation}
where $\bm n_E$ is the outward unit normal to $\partial E$, and
\begin{align*}
\begin{split}
\bm V=&\left\{\bm v\in L^2(\Omega)^2: \bm v|_{E}\in (H^1(E))^2, \forall E\in\mathscr{E}_h\right\},\\
\bm V^2=&\left\{\bar{\bm\sigma}\in (L^2(\Omega)^2)^2 :\bar{\bm\sigma}|_E\in((H^1(E))^2)^2 , \forall E\in\mathscr{E}_h\right\},\\
M=&\left\{q\in L^2(\Omega) : \int_{\Omega}qd\bm x=0 , q|_E\in H^1(E), \forall E\in\mathscr{E}_h\right\}.
\end{split}
\end{align*}
The above equations are well defined for any functions $(\bm u, \bar{\bm\sigma}, p)$ and $(\bm v, \bar{\bm\tau}, q)$ belonging to $\bm V\times\bm V^2\times M$.

The exact solution $(\bm u,\bar{\bm\sigma}, p)$ will be approximated by the functions $(\bm u_h,\bar{\bm\sigma}_h, p_h)$ belonging to the finite element spaces $\bm V_h\times\bm V^2_h\times M_h$ :
\begin{align*}
\begin{split}
\bm V_h&=\left\{\bm v\in L^2(\Omega)^2: \bm v|_{E}\in (\mathbb{P}^k(E))^2 , \forall E\in\mathscr{E}_h\right\} ,\\
\bm V^2_h&=\left\{\bar{\bm\sigma}\in (L^2(\Omega)^2)^2 :\bar{\bm\sigma}|_E\in((\mathbb{P}^{k}(E))^2)^2 , \forall E\in\mathscr{E}_h\right\} ,\\
M_h&=\left\{q\in L^2(\Omega) : \int_{\Omega}qd\bm x=0 , q|_E\in
\mathbb{P}^k (E) , \forall E\in\mathscr{E}_h\right\};
  \end{split}
\end{align*}
and the space $\tilde{\bm V}_h =\left\{\bm v_h\in\bm V_h: \forall e\in\mathscr{E}_h^B, \lbrack\bm v_h\rbrack|_e\cdot\bm n_e=0 \right\}$ will also be used in the following analysis.
That is to find $(\bm u_h,\bar{\bm\sigma}_h, p_h) \in \bm V_h\times\bm V^2_h\times M_h$ such that for any $(\bm v, \bar{\bm \tau}, q) \in \bm V_h\times\bm V^2_h\times M_h$ and $E \in \mathscr{E}_h$, the following holds:
\begin{equation} \label{2.5}
\left\{ \begin{array}
 {l@{\quad} l}
 \int_E((\bm u_h)_t+(\bm u_h\cdot\nabla)\bm u_h)\cdot\bm v+\int_E\sqrt{\nu}\bar{\bm\sigma}_h:\nabla\bm v-\int_{\partial E}\sqrt{\nu}\bar{\bm\sigma}_h^{\ast}\cdot\bm v\cdot\bm n_E\\
-\int_E p_h\nabla\cdot\bm v+\int_{\partial E}p_h^{\ast}\bm v\cdot \bm n_E=\int_E\bm f\cdot\bm v,\ \ \ \ \ \ \forall \bm v\in\bm V_h,\\
\\
\int_E\bar{\bm\sigma}_h:\bar{\bm\tau}-\int_E\sqrt{\nu}\nabla\bm u_h:\bar{\bm\tau}=0,\ \ \ \ \ \ \ \ \ \ \ \ \ \ \ \ \forall \bar{\bm\tau}\in\bm V_h^2,\\
\\
\int_E\nabla\cdot\bm u_h q=0,\ \ \ \ \ \ \ \ \ \ \ \ \ \ \ \ \ \ \ \ \ \ \ \ \ \ \ \ \ \ \ \ \ \ \ \forall q\in M_h,
 \end{array}
 \right.
\end{equation}
where $\bar{\bm\sigma}_h^{\ast}$ and $p_h^{\ast}$ are to be determined numerical fluxes. By carefully adding the penalty terms and choosing the numerical fluxes:
\begin{align}
\bar{\bm\sigma}^{\ast}_h=\{\{\bar{\bm\sigma}_h\}\},\ \ \ \ \ \ \ \ \ p^{\ast}_h=\{\{p_h\}\};
\end{align}
%and if $e$ lies on the boundary
%\begin{align}
%\bar{\bm\sigma}^{\ast}=\bar{\sigma},\ \ \ \ \ \ \ \   \ \ \  p^{\ast}=p;
%\end{align}
we develop the following numerical scheme:
\begin{equation} \label{2.7}
\left\{ \begin{array}
 {l@{\quad} l}
  \sum_{E\in\mathscr{E}_h}( (\bm u_h)_t+(\bm u_h\cdot\nabla)\bm u_h,\bm v)_E+\sum_{E\in\mathscr{E}_h}(\bar{\bm\sigma}_h,\sqrt{\nu}\nabla\bm v)_E\\-\sum_{e\in\mathscr{E}_h^B}(\{\{\bar{\bm\sigma}_h\}\},\sqrt{\nu}\lbrack\bm v\rbrack\otimes\bm n_e)_e
 -\sum_{ E\in\mathscr{E}_h}(p_h,\nabla\cdot\bm v)_E\\
+\sum_{e\in\mathscr{E}_h^B}(\{\{p_h\}\},\lbrack\bm v\rbrack\cdot\bm n_e)_e=\sum_{E\in\mathscr{E}_h}(\bm f,\bm v)_E, \\
\\
 %\cr\noalign{\vskip  0.01 mm}
 \sum_{E\in\mathscr{E}_h}(\bar{\bm\sigma}_h,\bar{\bm\tau})_E
-\sum_{E\in\mathscr{E}_h}(\sqrt{\nu}\nabla\bm u_h,\bar{\bm\tau})_E\\
+\sum_{e\in\mathscr{E}_h^B}(\{\{\bar{\bm\tau}\}\},\sqrt{\nu}\lbrack\bm u_h\rbrack\otimes\bm n_e)_e=0, \\
\\
 \sum_{E\in\mathscr{E}_h}(q,\nabla\cdot\bm u_h)_E-
\sum_{e\in\mathscr{E}_h^B}(\{\{q\}\},\lbrack\bm u_h\rbrack\cdot\bm n_e)_e=0,
 \end{array}
 \right.
\end{equation}
for any $(\bm v, \bar{\bm \tau}, q) \in \bm V_h\times\bm V^2_h\times
M_h$. The exact solution of (\ref{1.1}) is expected to be at least
continuous; and the boundary is homogeneous; so the added penalty
terms
$\sum_{e\in\mathscr{E}_h^B}(\{\{\bar{\bm\tau}\}\},\sqrt{\nu}\lbrack\bm
u_h\rbrack\otimes\bm n_e)_e$ and
$\sum_{e\in\mathscr{E}_h^B}(\{\{q\}\},\lbrack\bm u_h\rbrack\cdot\bm
n_e)_e$ still keep the consistency of the scheme. Moreover, the
locality of the discontinuous Galerkin method still remains since
the penalty in the second equation is about $\bm u_h$
element-by-element and it is independent of    $\bar{\bm\sigma}_h$.
The most important thing is that these two additions make the
variational formulation well-symmetric. Then it makes the
theoretical analysis and numerical implementation of the scheme
convenient.

%Note that the penalty in the second equation is about $\bm u_h$ element-by-element. Since it is independent of
%   $\bar{\bm\sigma}$, we can eliminate the term $\bar{\bm\sigma}_h$ from the equations.

%\begin{itemize}
%  \item  Note that the penalty in the second equation is about $\bm u$ element-by-element. Since it is independent of
%   $\bar{\bm\sigma}$, we can eliminate the term $\bar{\bm\sigma}$ from the equations.
%
%\item Noticeably , the fluxes are consistent in equations (2.7).
%The fluxes in the first equation of (2.7) is classical central fluxes. The penalty in the second and the third equations of (2.7) are intrinsical zero. Since the boundary condition $\bm u=0$ and the continuous quality, we have  $\forall e\in\mathscr{E}_h^B ,\ \ \lbrack\bm u\rbrack|_e=0$.
%  \item  Although the method needs some boundary conditions, the advantage is obvious.
%  The fluxes we choose are easy and the form is simple compared with the IIPG , SIPG , NIPG \cite{r12}. The importance of this weak variational form is easily understand and implemented .
%  \item  Since $\bar{\bm\sigma}=\sqrt{\nu}\nabla\bm u$, we find if the viscosity coefficient $\nu$ is rather small ,
%  like $1/100000000$,and $\sqrt{\nu}=1/10000$ is much bigger than $\nu$,  the discrete matrix is still stable. This is an amazing discovery.
%\end{itemize}

\vskip 0.2cm

\textbf{Definitions of the bilinear form:}

\begin{align*}
\begin{split}
a(\bar{\bm\sigma}_h,\bm v)&=\sum_{E\in\mathscr{E}_h}(\bar{\bm\sigma}_h,\sqrt{\nu}\nabla\bm v)_E-\sum_{e\in\mathscr{E}_h^B}
(\{\{\bar{\bm\sigma}_h\}\},\sqrt{\nu}\lbrack\bm v\rbrack\otimes\bm n_e)_e, \\
b(p_h,\bm v)&=-\sum_{E\in\mathscr{E}_h}(p_h,\nabla\cdot\bm v)_E+
\sum_{e\in\mathscr{E}_h^B}(\{\{p_h\}\},\lbrack\bm v\rbrack\cdot\bm n_e)_e, \\
A(\bar{\bm\sigma}_h,\bar{\bm\tau})&=\sum_{E\in\mathscr{E}_h}(\bar{\bm\sigma}_h,\bar{\bm\tau})_E.
\end{split}
\end{align*}
Then the numerical scheme (\ref{2.7}) can be recast as for any $(\bm v, \bar{\bm \tau}, q) \in \bm V_h\times\bm V^2_h\times M_h$:
\begin{equation}
\left\{ \begin{array}
 {l@{\quad} l}
 \sum_{E\in\mathscr{E}_h}((\bm u_h)_t+(\bm u_h\cdot\nabla)\bm u_h,\bm v)_E+
a(\bar{\bm\sigma}_h,\bm v)+b(p_h,\bm v)=(\bm f,\bm v), \\ \\
 %\cr\noalign{\vskip  0.01 mm}
  A(\bar{\bm\sigma}_h,\bar{\bm\tau})-a(\bar{\bm\tau},\bm u_h)=0, \\ \\
  -b(q,\bm u_h)=0.
 \end{array}
 \right.
\end{equation}
%Thus,we see that (2.8) is symmetric (after changing a sign in equations). This is in sharp contrast with the
%original Equation of (1.1), which is nonsymmetric due to the presence of the advection term.

%\vskip 0.2cm
%
%\textbf{Discrete inf-sup condition:}
%
%\vskip 0.2cm
%
%\textbf{Lemma 2.4} (Div-Grad relation\cite{r21}). For all $\bm v\in\tilde{\bm V}_h\subset H_0^1(\Omega)^2$, the following inequality holds:
%\begin{equation}
%\parallel\nabla\cdot\bm v\parallel_{L^2(\Omega)}\leq C\parallel\nabla\bm v\parallel_{(L^2(\Omega)^2)^2}.
%\end{equation}
%
%\textbf{Lemma 2.5.} The Raviart-Thomas interpolation\cite{r15} $\pi : H^1(\Omega)^2 \longrightarrow \bm V_h$ satisfies for all $\bm v\in H^1(\Omega)^2$
% \begin{equation}
% \forall E\in\mathscr{E}_h\ ,\ \forall q\in \mathbb{P}^{k-1}(E),\ \int_{E}q\nabla\cdot(\pi\bm v-\bm v)=0\ ,
% \end{equation}
% \begin{equation}
% \forall e\in\mathscr{E}_h^B\ ,\ \forall q\in \mathbb{P}^{k-1}(e),\ \int_{e}q(\pi\bm v-\bm v)\cdot\bm n_e=0\ ,
% \end{equation}
% \begin{equation}
% \forall e\in\mathscr{E}_h^B\ ,\ \pi\bm v|_e\cdot\bm n_e\in \mathbb{P}^{k-1}(e)\ ,
% \end{equation}
% \begin{equation}
% \forall E\in\mathscr{E}_h\ ,\parallel\pi\bm v-\bm v\parallel_{L^2(E)^2}+h_E\parallel\nabla(\pi\bm v-\bm v)\parallel_{(L^2(E)^2)^2}\leq C h_E\parallel\nabla\bm v\parallel_{(L^2(E)^2)^2}\ ,
% \end{equation}
% \begin{equation}
% \parallel\pi\bm v\parallel_{\varepsilon_1}\leq C\parallel\nabla\bm v\parallel_{(L^2(E)^2)^2}\ .
%\end{equation}
\vskip 0.2cm
\textbf{Lemma 2.1} (Discrete Inf-Sup\cite{r15}). There exists a constant $\beta^*> 0$, independent of $h$, such that
\begin{equation} \label{2.15}
\inf_{q\in M_h}\sup_{\bm v\in \tilde{\bm V}_h}\frac{b(q, \bm v)}{\parallel\bm v\parallel_{\varepsilon_1}\parallel q\parallel_{L^2(\Omega)}}\geq \beta^*,
\end{equation}
where
$$
 %b(\bm v,q)&=-\sum_{E\in\mathscr{E}_h}\int_{E}q\nabla\cdot\bm v+\sum_{e\in\mathscr{E}_h^B}\int_e\{\{q\}\}\lbrack\bm v\rbrack\cdot\bm n_e \ ,\\
 \parallel\bm v\parallel_{\varepsilon_1}=\Big(\sum_{E\in\mathscr{E}_h}\parallel\nabla\bm v\parallel^2_{(L^2(E)^2)^2}+\sum_{e\in\mathscr{E}_h^B}\parallel\lbrack\bm v\rbrack\parallel_{L^2(e)^2}^2\Big)^{1/2} .
$$

\subsection{\textbf{Characteristics method}}

For each positive integer $N$, let $0=t^0<t^1<\cdots<\cdots<t^N=T$ be a partition of $T$ into subintervals
$T^n=(t^{n-1},t^n]$, with uniform mesh and the interval length $\Delta t=t^n-t^{n-1},1\leq n\leq N$. And denote $\bm u^n=\bm u(\bm x,t^n)$. The characteristics tracing back along the field of a point $\bm x\in\Omega$ at time $t^n$ to $t^{n-1}$ is approximately\cite{r1,r2,r3}
$$\check{\bf x}(\bm x,t^{n-1})=\bm x-\bm u^{n-1}\Delta t. $$
For the discontinuous Galerkin method, $\forall\  \bm x\in E$, we
must have $\check{\bf x}(\bm x,t^{n-1})=\bm x-\bm u^{n-1}\Delta t\in
E$ which implies that the $\Delta t$ must be small enough to ensure
the property. If the initial value is rather small, it can also
ensure the property. Consequently, the approximation for the
hyperbolic part of (1.1) at time $t^n$ can be derived as follows:
$$\bm u_t^n+\bm u^{n}\cdot\nabla\bm u^n=\bm u_t^n+\bm u^{n-1}\cdot\nabla\bm u^{n}+\mathcal{O}(\Delta t);$$
$$ \bm u_t^n=\frac{\bm u^n-\bm u^{n-1}}{\Delta t}+\mathcal{O}(\Delta t);~~{\rm and ~denote} ~~\check{\bm u}^{n-1}=\bm u(\check{\bf x},t^{n-1});$$
and
\begin{align*}
\bm u(\bm x-\bm u^{n-1}\Delta t,t^{n-1})
=\bm u(\bm x,t^{n-1})-\bm u^{n-1}\Delta t\cdot\nabla\bm u^{n-1}+(\bm u^{n-1}\Delta t)^2\Delta\bm u^{n-1}/2!+\cdots.
\end{align*}
Then there exists
$$\bm u^{n-1}\cdot\nabla\bm u^{n-1}=\frac{\bm u^{n-1}-\check{\bm u}^{n-1}}{\Delta t}+\mathcal{O}(\triangle t).$$
From all above, we get the characteristic method:
$$\bm u_t^n+\bm u^{n-1}\cdot\nabla \bm u^{n}=\frac{\bm u^n-\check{\bm u}^{n-1}}{\triangle t}+\mathcal{O}(\triangle t).$$
%On a triangulation, we define two finite dimensional subspaces $\bm X_h\subset\bm X$ and $M_h\subset M$:
%$$\bm X_h=\left\{\bm v_h\in(L^2(\Omega))^2:\forall E\in\mathscr{E}_h;\bm v_h\in(P_k(E))^2\right\} ,$$
%$$M_h=\left\{q_h\in(L^2_0(\Omega)):\forall E\in\mathscr{E}_h;q_h\in(P_k(E))\right\} .$$
%where $k$ is the degree of approximated polynomials.

So the fully discretized scheme, i.e., the characteristic local
discontinuous Galerkin (CLDG) scheme, corresponding to the
variational formulation (\ref{2.7}) is to find $(\bm u_h^n,
\bar{\bm\sigma}_h^n, p_h^n) \in \bm V_h\times\bm V_h^2\times M_h$
such that
\begin{subequations}
\begin{equation}\label{2.16a}
\begin{split}
  \Big(\frac{\bm u_h^n-\check{\bm u}_h^{n-1}}{\triangle t},\bm v\Big)+\sum_{E\in\mathscr{E}_h}(\bar{\bm\sigma}_h^n,\sqrt{\nu}\nabla\bm v)_E
-\sum_{e\in\mathscr{E}^B_h}(\{\{\bar{\bm\sigma}^n_h\}\},\sqrt{\nu}\lbrack\bm v\rbrack\otimes\bm n_e)_e\\
-\sum_{E\in\mathscr{E}_h}(p_h^n,\nabla\cdot\bm v)_E
+\sum_{e\in\mathscr{E}_h^B}(\{\{p_h^n\}\},\lbrack\bm v\rbrack\cdot\bm n_e)_e=(\bm f^n,\bm v),\ \ \ \   \forall \ \bm v\in\bm V_h,
\end{split}
\end{equation}\label{2.16b}
\begin{equation}
\begin{split}
 \sum_{E\in\mathscr{E}_h}(\bar{\bm\sigma}^n_h,\bar{\bm\tau})_E-\sum_{E\in\mathscr{E}_h}(\sqrt{\nu}\nabla\bm u_h^n,\bar{\bm\tau})_E
 +\sum_{e\in\mathscr{E}_h^B}(\{\{\bar{\bm\tau}\}\},\sqrt{\nu}\lbrack\bm u_h^n\rbrack\otimes\bm n_e)_e=0,\\
 \ \ \ \ \ \ \ \ \ \ \ \ \ \ \ \forall\ \bar{\bm\tau}\in\bm V_h^2,
 \end{split}
\end{equation}
\begin{equation}\label{2.16c}
\ \ \ \ \ \ \ \  \ \ \ \ \  \ \ \ \ \ \ \ \ \  \sum_{E\in\mathscr{E}_h}(q,\nabla\cdot\bm u_h^n)_E-\sum_{e\in\mathscr{E}_h^B}(\{\{q\}\},\lbrack\bm u_h^n\rbrack\cdot\bm n_e)_e=0,\ \ \ \forall\ q\in M_h .
\end{equation}
\end{subequations}

\section{Existence, uniqueness, and stability analysis}

\subsection{Existence and uniqueness}

For the notational convenience we define the following equation:
\begin{equation}
\begin{split}
&\ \ \ \ \  \mathscr{A}(\bm u_h^n,\bar{\bm\sigma}_h^n,p_h^n;\bm v,\bar{\bm\tau},q)\\
&=\sum_{E\in\mathscr{E}_h}(\bar{\bm\sigma}_h^n,\sqrt{\nu}\nabla\bm v)_E-\sum_{e\in\mathscr{E}_h^B}(\{\{\bar{\bm\sigma}_h^n\}\},\sqrt{\nu}\lbrack\bm v\rbrack\otimes\bm n_e)_e
\\
&~~ -\sum_{E\in\mathscr{E}_h}(p_h^n,\nabla\cdot\bm v)+\sum_{e\in\mathscr{E}_h^B}(\{\{p_h^n\}\},\lbrack\bm v\rbrack\cdot\bm n_e)_e
\\
&~~ +\sum_{E\in\mathscr{E}_h}(\bar{\bm\sigma}_h^n,\bar{\bm\tau})_E-\sum_{E\in\mathscr{E}_h}(\sqrt{\nu}\nabla\bm u_h^n,\bar{\bm\tau})_E
\\
&~~ +\sum_{e\in\mathscr{E}_h^B}(\{\{\bar{\bm\tau}\}\},\sqrt{\nu}\lbrack\bm u_h^n\rbrack\otimes\bm n_e)_e
+\sum_{E\in\mathscr{E}_h}(q,\nabla\cdot\bm u_h^n)_E\\
&~~ -\sum_{e\in\mathscr{E}_h^B}(\{\{q\}\},\lbrack\bm u_h^n\rbrack\cdot\bm n_e)_e;
\end{split}
\end{equation}
and the right side hand
\begin{equation}
\mathscr{F}(\bm v)=(\bm f^n,\bm v).
\end{equation}
Hence, the linear system of equations (\ref{2.16a})-(\ref{2.16c}) can be written equivalently as:
\begin{align}
\begin{split}\label{3.3}
&\ \ \ \ \  \frac{1}{\Delta t}(\bm u_h^n,\bm v)+\mathscr{A}(\bm u_h^n,\bar{\bm\sigma}_h^n,p_h^n;\bm v,\bar{\bm\tau},q)\\
&=\mathscr{F}(\bm v)+\frac{1}{\Delta t}(\check{\bm u}^{n-1}_h,v),\,\,\,\forall\ (\bm v,\bar{\bm\tau},q)\in\bm V_h\times\bm V_h^2\times M_h.
\end{split}
\end{align}
\textbf{Lemma 3.1.} There exists a unique solution $(\bm u_h^n,\bar{\bm\sigma}_h^n,p_h^n)\in\bm V_h\times\bm V_h^2\times M_h$ satisfying (3.3).
\\
\\
{\bf Proof}: To ensure the computability of the algorithm, we begin by showing that the variational formulation (\ref{3.3}) is uniquely solvable for
$(\bm u_h^n,\bar{\bm\sigma}_h^n,p_h^n)$ at each time step $n$. As (\ref{3.3}) represents a finite system of linear
equations, the uniqueness of $(\bm u_h^n, \bar{\bm\sigma}_h^n,p_h^n)$ is equivalent to the existence.
\\
Letting $\check{\bm u}_h^{n-1}=\bm f=0$ and taking $\bm v=\bm u_h^n, \bar{\bm\tau}=\bar{\bm\sigma}_h^n, q=p_h^n$, we have
\begin{align*}
\begin{split}
&\ \ \ \ \frac{1}{\Delta t}(\bm u_h^n,\bm u_h^n)+\mathscr{A}(\bm u_h^n,\bar{\bm\sigma}_h^n,p_h^n;\bm u_h^n, \bar{\bm\sigma}_h^n,p_h^n) \\
&=\frac{1}{\triangle t}(\bm u_h^n,\bm u_h^n)+\sum_{E\in\mathscr{E}_h}(\bar{\bm\sigma}_h^n,\sqrt{\nu}\nabla\bm u_h^n)_E\\
&-\sum_{e\in\mathscr{E}_h^B}(\{\{\bar{\bm\sigma}_h^n\}\},\sqrt{\nu}\lbrack\bm u_h^n\rbrack\otimes\bm n_e)_e
-\sum_{E\in\mathscr{E}_h}(p_h^n,\nabla\cdot\bm u_h^n)
\\
&+\sum_{e\in\mathscr{E}_h^B}(\{\{p_h^n\}\},\lbrack\bm u_h^n\rbrack\cdot\bm n_e)_e
+\sum_{E\in\mathscr{E}_h}(\bar{\bm\sigma}_h^n-\sqrt{\nu}\nabla\bm u_h^n,\bar{\bm\sigma}_h^n)_E\\
&+\sum_{e\in\mathscr{E}_h^B}(\{\{\bar{\bm\sigma}_h^n\}\},\sqrt{\nu}\lbrack\bm u_h^n\rbrack\otimes\bm n_e)_e
+\sum_{E\in\mathscr{E}_h}(p_h^n,\nabla\cdot\bm u_h^n)_E\\
&-\sum_{e\in\mathscr{E}_h^B}(\{\{p_h^n\}\},\lbrack\bm u_h^n\rbrack\cdot\bm n_e)_e.
\end{split}
\end{align*}
Therefore,
\begin{align}
\begin{split}
&\ \ \ \ \ \frac{1}{\Delta t}(\bm u_h^n,\bm u_h^n)+\mathscr{A}(\bm u_h^n,\bar{\bm\sigma}_h^n,p_h^n;\bm u_h^n, \bar{\bm\sigma}_h^n,p_h^n)\\
&=\frac{1}{\triangle t}(\bm u_h^n,\bm u_h^n)+\sum_{E\in\mathscr{E}_h}(\bar{\bm\sigma}_h^n,\bar{\bm\sigma}_h^n)_E\geq 0
\end{split}
\end{align}
Further letting $\bm u_h^n=\bar{\bm\sigma}_h^n=0$,  for $p_h^n \in M_h$, there exists
$$\forall \bm v\in\bm V_h,\ \ \ b(\bm v,p_h^n)=0;$$
and from Lemma 2.1, we get
$$\parallel p^n_h\parallel_{L^2(\Omega)}\leq \sup_{\bm v\in\tilde{\bm V}_h}\frac{b(\bm v,p_h^n)}{\beta^*\parallel\bm v\parallel_{\varepsilon_1}}=0.$$
Hence, $p_h^n=0$, which completes the proof of the uniqueness of the solution.
 \\
 \vskip 0.1cm

\noindent\textbf{Remark}. Here $\bm u_h^n$ and $p_h^n$ can be solved simultaneously, being different from some of the other methods which need to first solve  $\bm u_h^n$ then $p_h^n$.

\subsection{Stability analysis}
In this subsection, we present and prove the numerical stability result.
 \vskip 0.1cm
\noindent \textbf{Theorem 3.2} (Numerical stability). The CLDG scheme (\ref{3.3}) is unconditionally stable, i.e., for any integer $N=1,2,3,\cdots$, there exists
 \begin{align*}
\begin{split}
& \parallel\bm u_h^N\parallel_{L^2(\mathscr{E}_h)^2}^2+2\Delta t\sum_{n=1}^N\parallel\bar{\bm\sigma}_h^n\parallel_{({L^2(\mathscr{E}_h)^2})^2}^2 \\
& \leq e^{CT} \left(
  \Delta t\sum_{n=1}^N\parallel\bm f^n\parallel_{L^2(\mathscr{E}_h)^2}^2+(C\Delta t+1)\parallel\bm u_h^0\parallel_{L^2(\mathscr{E}_h)^2}^2 \right),
\end{split}
\end{align*}
 where $\bm u_h^0=\bm u^0$, and $C$ is a constant depending on $\nabla \bm u$.
\vskip 0.1cm
\noindent{\bf Proof}: Taking $\bm v=2\Delta t\bm u_h^n$, $\bar{\bm\tau}=\bar{\bm\sigma}_h^n$, and $q=p_h^n $, respectively, in (\ref{2.16a})-(\ref{2.16c}), we get the following equations:
\begin{equation}\label{3.5}
\begin{split}
 &\ \ \ \ \big(\frac{\bm u_h^n-
\check{\bm u}_h^{n-1}}{\Delta t},2\Delta t\bm u_h^n\big)+\sum_{E\in\mathscr{E}_h}(\bar{\bm\sigma}_h^n,2\Delta t\sqrt{\nu}\nabla\bm u_h^n)_{E}\\
&-\sum_{e\in\mathscr{E}_h^B}(\{\{\bar{\bm\sigma}_h^n\}\},2\Delta t\sqrt{\nu}\lbrack\bm u_h^n\rbrack\otimes\bm n_e)_e
-\sum_{E\in\mathscr{E}_h}(p_h^n,2\Delta t\nabla\cdot\bm u_h^n)_E\\
&+\sum_{e\in\mathscr{E}_h^B}(\{\{p_h^n\}\},2\Delta t\lbrack\bm u_h^n\rbrack\cdot\bm n_e)_e=(\bm f^n,2\Delta t\bm u_h^n),
\end{split}
\end{equation}
\begin{equation}\label{3.6}
\sum_{E\in\mathscr{E}_h}(\bar{\bm\sigma}_h^n,\bar{\bm\sigma}_h^n)_E-\sum_{E\in\mathscr{E}_h}(\sqrt{\nu}\nabla\bm u_h^n,\bar{\bm\sigma}_h^n)_E+\sum_{e\in\mathscr{E}_h^B}(\{\{\bar{\bm\sigma}_h^n\}\},\sqrt{\nu}\lbrack\bm u_h^n\rbrack\otimes\bm n_e)_e=0,
\end{equation}
and
\begin{equation}\label{3.7}
\ \ \ \ \ \ \ \ \ \ \ \ \ \ \ \ \ \ \ \ \ \ \ \ \ \ \ \ \ \ \ \ \ \ \ \ \ \ \ \sum_{E\in\mathscr{E}_h}(p_h^n,\nabla\cdot\bm u_h^n)_E-\sum_{e\in\mathscr{E}_h^B}(\{\{p_h^n\}\},\lbrack\bm u_h^n\rbrack\cdot\bm n_e)_e=0.
\end{equation}
Multiplying $2\Delta t$ in (\ref{3.6}), $2\Delta t$ in (\ref{3.7}), and adding all the above equations lead to
\begin{align*}
\begin{split}
&\ \ \ \ \ (\bm f^n,2\Delta t\bm u_h^n)\\
&=2(\bm u_h^n-\check{\bm u}_h^{n-1},\bm u_h^n)+\sum_{E\in\mathscr{E}_h}(\bar{\bm\sigma}_h^n,2\Delta t\sqrt{\nu}\nabla\bm u_h^n)_E\\
&-\sum_{e\in\mathscr{E}_h^B}(\{\{\bar{\bm\sigma}^n_h\}\},2\Delta t\sqrt{\nu}\lbrack\bm u_h^n\rbrack\otimes\bm n_e)_e
-\sum_{E\in\mathscr{E}_h}(p_h^n,2\Delta t\nabla\cdot\bm u_h^n)_E\\
&+\sum_{e\in\mathscr{E}_h^B}(\{\{p_h^n\}\},2\Delta t\lbrack\bm u_h^n\rbrack\cdot\bm n_e)_e+\sum_{E\in\mathscr{E}_h}2\Delta t(\bar{\bm\sigma}_h^n,\bar{\bm\sigma}_h^n)_E\\
&-\sum_{E\in\mathscr{E}_h}(2\Delta t\sqrt{\nu}\nabla\bm u_h^n,\bar{\bm\sigma}_h^n)_E+\sum_{e\in\mathscr{E}_h^B}(\{\{\bar{\bm\sigma}_h^n\}\},2\Delta t\sqrt{\nu}\lbrack\bm u_h^n\rbrack\otimes\bm n_e)_e\\
&+\sum_{E\in\mathscr{E}_h}(p_h^n,2\Delta t\nabla\cdot\bm u_h^n)_E-\sum_{e\in
\mathscr{E}_h^B}(\{\{p_h^n\}\},2\Delta t\lbrack\bm u_h^n\rbrack\cdot\bm n_e)_e.
\end{split}
\end{align*}
Then, we get
$$2(\bm u_h^n-\check{\bm u}_h^{n-1},\bm u_h^n)+2\Delta t(\bar{\bm\sigma}_h^n,\bar{\bm\sigma}_h^n)=2\Delta t(\bm f^n,\bm u_h^n)\ .$$
Since
$$
\begin{array}{ll}
2(\bm u_h^n-\check{\bm u}^{n-1}_h,\bm u_h^n)&=\parallel\bm u_h^n\parallel_{L^2(\mathscr{E}_h)^2}^2-\parallel\check{\bm u}^{n-1}_h\parallel_{L^2(\mathscr{E}_h)^2}^2+\parallel\bm u_h^n-\check{\bm u}_h^{n-1}\parallel_{L^2(\mathscr{E}_h)^2}^2
\\
&\geq\parallel\bm u_h^n\parallel_{L^2(\mathscr{E}_h)^2}^2-\parallel\check{\bm u}_h^{n-1}\parallel_{L^2(\mathscr{E}_h)^2}^2;
\end{array}
$$
and
\begin{align}
\begin{split}
&\parallel\check{\bm u}_h^{n-1}\parallel_{L^2(\mathscr{E}_h)^2}^2-\parallel\bm u_h^{n-1}\parallel_{L^2(\mathscr{E}_h)^2}^2  \\
&= \int_{\mathscr{E}_h}(\check{\bm u}_h^{n-1}\cdot \check{\bm u}_h^{n-1})d\bm x
-\int_{\mathscr{E}_h}(\bm u_h^{n-1}\cdot \bm u_h^{n-1})d\bm x \\
&=\int_{\mathscr{E}_h}(\bm u_h^{n-1}\cdot \bm u_h^{n-1})(1+\mathcal {O}(\Delta t))d\bm x-\int_{\mathscr{E}_h}(\bm u_h^{n-1} \cdot \bm u_h^{n-1})d\bm x \\
&\leq C\Delta t\parallel\bm u_h^{n-1}\parallel_{L^2(\mathscr{E}_h)^2}^2.
\end{split}
\end{align}
%$$\parallel\check{\bm u}_h^{n-1}\parallel_0^2-\parallel\bm u_h^{n-1}\parallel_0^2\leq C\Delta t\parallel\bm u_h^{n-1}\parallel_0^2.$$
Then, we obtain
\begin{align*}
\begin{split}
& \parallel\bm u_h^n\parallel_{L^2(\mathscr{E}_h)^2}^2-\parallel\bm u_h^{n-1}\parallel_{L^2(\mathscr{E}_h)^2}^2+2\Delta t\parallel\bar{\bm \sigma}_h^n\parallel_{(L^2(\mathscr{E}_h)^2)^2}^2 \\
& \leq C\Delta t\parallel\bm u_h^{n-1}\parallel_{L^2(\mathscr{E}_h)^2}^2+2\Delta t\parallel\bm f^n\parallel_{L^2(\mathscr{E}_h)^2} \parallel\bm u_h^n\parallel_{L^2(\mathscr{E}_h)^2}.
\end{split}
\end{align*}
Thus,
\begin{equation}
\begin{split}
& \parallel\bm u_h^n\parallel_{L^2(\mathscr{E}_h)^2}^2-\parallel\bm u_h^{n-1}\parallel_{L^2(\mathscr{E}_h)^2}^2
+2\Delta t\parallel\bar{\bm\sigma}_h^n\parallel_{(L^2(\mathscr{E}_h)^2)^2}^2\\
&\leq C\Delta t\parallel\bm u_h^{n-1}\parallel_{L^2(\mathscr{E}_h)^2}^2+\Delta t\parallel\bm f^n\parallel_{L^2(\mathscr{E}_h)^2}^2+\Delta t\parallel \bm u_h^{n}\parallel_{L^2(\mathscr{E}_h)^2}^2.
\end{split}
\end{equation}
Summing up the above equation from $n=1$ to $N$, we have
\begin{align*}
\begin{split}
& \parallel\bm u_h^N\parallel_{L^2(\mathscr{E}_h)^2}^2-\parallel\bm u_h^0\parallel_{L^2(\mathscr{E}_h)^2}^2+2\Delta t\sum_{n=1}^N\parallel\bar{\bm\sigma}_h^n\parallel_{(L^2(\mathscr{E}_h)^2)^2}^2
\\
&
\leq
C\Delta t\sum_{n=1}^N\parallel\bm u_h^{n-1}\parallel_{L^2(\mathscr{E}_h)^2}^2+\Delta t\sum_{n=1}^N\parallel\bm f^n\parallel_{L^2(\mathscr{E}_h)^2}^2+\Delta t\sum_{n=1}^N\parallel \bm u_h^n\parallel_{L^2(\mathscr{E}_h)^2}^2.
\end{split}
\end{align*}
Then the following holds.
\begin{align*}
\begin{split}
& \parallel\bm u_h^N\parallel^2_{L^2(\mathscr{E}_h)^2}+2\Delta t\sum_{n=1}^N\parallel\bar{\bm\sigma}_h^n\parallel^2_{({L^2(\mathscr{E}_h)^2})^2} \\
&
\leq C\Delta t\sum_{n=1}^N\parallel\bm u_h^n\parallel_{L^2(\mathscr{E}_h)^2}^2+\Delta t\sum_{n=1}^N\parallel\bm f^n\parallel_{L^2(\mathscr{E}_h)^2}^2+(C\Delta t+1)\parallel\bm u_h^0\parallel_{L^2(\mathscr{E}_h)^2}^2.
\end{split}
\end{align*}
From the discrete Gronwall's inequality, we have
\begin{align*}
\begin{split}
& \parallel\bm u_h^N\parallel_{L^2(\mathscr{E}_h)^2}^2+2\Delta t\sum_{n=1}^N\parallel\bar{\bm\sigma}_h^n\parallel_{({L^2(\mathscr{E}_h)^2})^2}^2 \\
& \leq e^{CT} \left(
  \Delta t\sum_{n=1}^N\parallel\bm f^n\parallel_{L^2(\mathscr{E}_h)^2}^2+(C\Delta t+1)\parallel\bm u_h^0\parallel_{L^2(\mathscr{E}_h)^2}^2 \right).
\end{split}
\end{align*}
The proof is completed.

\section{Numerical experiment}

We perform the extensive numerical experiments to show the powerfulness of the presented schemes; comparing the numerical solutions with the constructed analytical ones, we show that the optimal convergence orders are obtained for the presented numerical schemes with a wide range of Reynolds numbers; one of the striking benefits of the proposed numerical schemes is that with the refining of the meshes the conditional number of the matrix $A$ of the matrix equation $Ax=b$ corresponding to the numerical schemes does not increase.  Furthermore, with the specified initial and boundary conditions and the given source term, we simulate the contours of the velocity and pressure at different time $t$.

%In this Section, we give two test examples. By the first example, we  verify the accuracy of our scheme with  the exact velocity solution $\bm u$ and pressure solution $p$. At the second example, we will draw some contour figures of velocity and pressure to illustrate the efficient methods along with different time with only the
%right-hand side function $\bm f=(f_1,f_2)$ and initial and boundary conditions of solution $\bm u$, but not exact solutions.
\vskip 0.2cm

\noindent\textbf{Example 5.1.} Suppose that the domain is the unit square $\Omega =[0,1]\times[0,1]$ and the smooth solution is given by\cite{r12}:

\begin{equation}
\left\{ \begin{array}
 {l@{\quad} l}
 u_1(x,y,t)=(x^4-2x^3+x^2)(4y^3-6y^2+2y)t,\\
 %\cr\noalign{\vskip  0.01 mm}
 u_2(x,y,t)=-(4x^3-6x^2+2x)(y^4-2y^3+y^2)t,\\
 p(x,y,t)=0 .
 \end{array}
 \right.
\end{equation}
\\
Then the exact solution $\bm u$ has the homogeneous boundary value. And the forcing term $\bm f(x,y,t)$ is determined accordingly from (\ref{1.1}) for any given $\nu$. In the numerical simulations, we respectively take $Re=1/\nu=1.0e+002, 1.0e+003, 1.0e+005$, and $1.0e+008$. Computations are performed with the $(\mathbb{P}^k, \mathbb{P}^k, \mathbb{P}^k)$ finite element pair on uniform meshes with the stepsize $h$. With a wide range of Reynolds numbers, the numerical results in Tables 1-8 illustrate that the optimal convergence orders ($(k+1)$ order accuracy) for both velocity and pressure are obtained.

%We expect a rate of convergence in $L^2(J^n\times L^2(\Omega))$  of order $\mathcal {O} (\Delta t+h^{k+1})$, where $k=1,2,3$. We impose $\Delta t=0.0001$ to avoid dominance of time error. Tables 1,2,3,4,5,6,7,8 illustrate $(k+1)$ order accuracy for both velocity and pressure in space step h. This computational results are better than the theory in Section 4. Note that these results are also better than that given at the reference \cite{r12}. We found that our result is not only good for big Reynolds number $(Re=10^8)$ and also efficient for getting a higher convergent order in spatial space.  Since the cost of the computational memory is huge for Discontinuous  Galerkin method for solving Navier-Stokes equations in two dimensional space, we choose the following h step to show the property of h-convergence.  For the sake of simpleness and internal storage, we choose the grid $K=128,200,288,392,512,648,800,968,1152,1352,1568,1800$. The numerical results are better compared with ref.\cite{r12}.

%%%%%%%%%%%%%%%%%%%%%%%%%%%%%%%%%%%%%%%%%%%%%%%%%%%%%%%%%%
\begin{table}[ht]
  \centering
  \begin{tabular}{|c|c|c|c|c|c|c|c|c|}
    \hline
    % after \\: \hline or \cline{col1-col2} \cline{col3-col4}
$K$    &$N=1,error$ & $order$    &$N=2,error$ &$order$      &$N=3,error$ & $order$ \\\hline
$128 $ & 3.437e-007 &\      \ \  & 1.775e-008 &\ \     \ \  & 8.627e-010 &\ \     \ \  \\
$200 $ & 2.267e-007 &\ 1.86\ \  & 9.175e-009 &\ \ 2.97\ \  & 3.565e-010 &\ \ 3.96\ \  \\
$288 $ & 1.582e-007 &\ 1.99\ \  & 5.195e-009 &\ \ 3.13\ \  & 1.688e-010 &\ \ 4.10\ \  \\
$392 $ & 1.165e-007 &\ 1.98\ \  & 3.301e-009 &\ \ 2.94\ \  & 9.371e-011 &\ \ 3.82\ \  \\
$512 $ & 8.872e-008 &\ 2.03\ \  & 2.200e-009 &\ \ 3.04\ \  & 5.602e-011 &\ \ 3.85\ \  \\
$648 $ & 7.042e-008 &\ 1.96\ \  & 1.530e-009 &\ \ 3.08\ \  & 3.443e-011 &\ \ 4.13\ \  \\
$800 $ & 5.770e-008 &\ 1.89\ \  & 1.117e-009 &\ \ 2.99\ \  & 2.278e-011 &\ \ 3.92\ \  \\
$968 $ & 4.765e-008 &\ 2.02\ \  & 8.273e-010 &\ \ 3.15\ \  & 1.538e-011 &\ \ 4.12\ \  \\
$1152$ & 4.018e-008 &\ 1.95\ \  & 6.415e-010 &\ \ 2.92\ \  & 1.086e-011 &\ \ 4.00\ \  \\
$1352$ & 3.410e-008 &\ 2.05\ \  & 4.979e-010 &\ \ 3.17\ \  & 7.907e-012 &\ \ 3.96\ \  \\
$1568$ & 2.944e-008 &\ 1.98\ \  & 3.930e-010 &\ \ 3.19\ \  & 5.858e-012 &\ \ 4.05\ \  \\
$1800$ & 2.559e-008 &\ 2.03\ \  & 3.239e-010 &\ \ 2.80\ \  & 4.468e-012 &\ \ 3.93\ \  \\ \hline
  \end{tabular}
  \caption{Numerical errors and convergence orders, for $\parallel\bm u(T)-\bm u_h^T\parallel_{L^2(\Omega)}$ with $\Delta t=1.0*10^{-4}$, $0005$, $Re=100$, i.e., $\nu=0.01$; $N$ is the degree of polynomial.}
  \end{table}
\begin{table}[ht]
  \centering
  \begin{tabular}{|c|c|c|c|c|c|c|c|c|}
    \hline
    % after \\: \hline or \cline{col1-col2} \cline{col3-col4}
$K$    &$N=1,error$ &$order$      &$N=2,error$&$order$     &$N=3,error$&$order$\\\hline
$128 $ & 3.439e-007 &\       \ \  &1.782e-008&\ \     \ \  &8.721e-010 &\ \     \ \  \\
$200 $ & 2.269e-007 &\  1.86 \ \  &9.236e-009&\ \ 2.95\ \  &3.616e-010 &\ \ 3.95\ \  \\
$288 $ & 1.584e-007 &\  1.97 \ \  &5.245e-009&\ \ 3.10\ \  &1.721e-010 &\ \ 4.07\ \  \\
$392 $ & 1.167e-007 &\  1.98 \ \  &3.342e-009&\ \ 2.92\ \  &9.567e-011 &\ \ 3.81\ \  \\
$512 $ & 8.890e-008 &\  2.04 \ \  &2.231e-009&\ \ 3.03\ \  &5.723e-011 &\ \ 3.85\ \  \\
$648 $ & 7.059e-008 &\  1.96 \ \  &1.561e-009&\ \ 3.03\ \  &3.533e-011 &\ \ 4.10\ \  \\
$800 $ & 5.787e-008 &\  1.88 \ \  &1.145e-009&\ \ 2.94\ \  &2.350e-011 &\ \ 3.87\ \  \\
$968 $ & 4.779e-008 &\  2.01 \ \  &8.515e-010&\ \ 3.11\ \  &1.587e-011 &\ \ 4.12\ \  \\
$1152$ & 4.034e-008 &\  1.95 \ \  &6.635e-010&\ \ 2.87\ \  &1.124e-011 &\ \ 3.96\ \  \\
$1352$ & 3.425e-008 &\  2.04 \ \  &5.172e-010&\ \ 3.11\ \  &8.169e-012 &\ \ 3.99\ \  \\
$1568$ & 2.958e-008 &\  1.98 \ \  &4.102e-010&\ \ 3.13\ \  &6.046e-012 &\ \ 4.06\ \  \\
$1800$ & 2.573e-008 &\  2.02 \ \  &3.399e-010&\ \ 2.72\ \  &4.607e-012 &\ \ 5.75\ \  \\\hline
  \end{tabular}
  \caption{Numerical errors and convergence orders, for $\parallel\bm u(T)-\bm u_h^T\parallel_{L^2(\Omega)}$ with $\Delta t=10^{-4}$ and $0005$, $Re=1000$, i.e., $\nu=0.001$; $N$ is the degree of polynomial.}
  \end{table}
  \begin{table}[ht]
  \centering
  \begin{tabular}{|c|c|c|c|c|c|c|c|c|}
    \hline
    % after \\: \hline or \cline{col1-col2} \cline{col3-col4}
$K$    &$N=1,error$&$order$    &$N=2,error$&$order$      &$N=3,error$&$order$ \\\hline
$128$  & 3.440e-007&\      \ \ &1.783e-008 &\ \     \ \  &8.732e-010&\ \    \ \  \\
$200$  & 2.269e-007&\  1.86\ \ &9.243e-009 &\ \ 2.94\ \  &3.622e-010&\ \ 3.94\ \  \\
$288$  & 1.584e-007&\  1.97\ \ &5.250e-009 &\ \ 3.10\ \  &1.725e-010&\ \ 4.07\ \  \\
$392$  & 1.168e-007&\  1.98\ \ &3.346e-009 &\ \ 2.92\ \  &9.593e-011&\ \ 3.81\ \  \\
$512$  & 8.892e-008&\  2.04\ \ &2.235e-009 &\ \ 3.02\ \  &5.740e-011&\ \ 3.85\ \  \\
$648$  & 7.061e-008&\  1.96\ \ &1.564e-009 &\ \ 3.03\ \  &3.547e-011&\ \ 4.09\ \  \\
$800$  & 5.789e-008&\  1.89\ \ &1.148e-009 &\ \ 2.93\ \  &2.361e-011&\ \ 3.86\ \  \\
$968$  & 4.781e-008&\  2.00\ \ &8.545e-010 &\ \ 3.10\ \  &1.596e-011&\ \ 4.11\ \  \\
$1152$ & 4.036e-008&\  1.95\ \ &6.663e-010 &\ \ 2.86\ \  &1.131e-011&\ \ 3.96\ \  \\
$1352$ & 3.427e-008&\  2.04\ \ &5.197e-010 &\ \ 3.10\ \  &8.226e-012&\ \ 3.98\ \  \\
$1568$ & 2.960e-008&\  1.98\ \ &4.125e-010 &\ \ 3.12\ \  &6.093e-012&\ \ 4.05\ \  \\
$1800$ & 2.575e-008&\  2.02\ \ &3.421e-010 &\ \ 2.71\ \  &4.648e-012&\ \ 3.92\ \  \\
 \hline
  \end{tabular}
  \caption{Numerical errors and convergence orders, for $\parallel\bm u(T)-\bm u_h^T\parallel_{L^2{\Omega}}$ with $\Delta t=10^{-4}$ and $0005$, $Re=50000$, i.e., $\nu=0.00002$; $N$ is the degree of polynomial.}
  \end{table}
\begin{table}[ht]
  \centering
  \begin{tabular}{|c|c|c|c|c|c|c|c|c|}
    \hline
    % after \\: \hline or \cline{col1-col2} \cline{col3-col4}
$K$    &$N=1,error$&$order$    &$N=2,error$&$order$     &$N=3,error$&$order$ \\\hline
$128 $ &3.440e-007&\      \ \  &1.783e-008&\ \     \ \  &8.733e-010&\ \     \ \  \\
$200 $ &2.269e-007&\  1.86\ \  &9.243e-009&\ \ 2.94\ \  &3.622e-010&\ \ 3.94\ \  \\
$288 $ &1.584e-007&\  1.97\ \  &5.250e-009&\ \ 3.10\ \  &1.725e-010&\ \ 4.07\ \  \\
$392 $ &1.168e-007&\  1.98\ \  &3.346e-009&\ \ 2.92\ \  &9.594e-011&\ \ 3.81\ \  \\
$512 $ &8.892e-008&\  2.04\ \  &2.235e-009&\ \ 3.02\ \  &5.740e-011&\ \ 3.85\ \  \\
$648 $ &7.061e-008&\  1.96\ \  &1.564e-009&\ \ 3.03\ \  &3.547e-011&\ \ 4.09\ \  \\
$800 $ &5.789e-008&\  1.89\ \  &1.148e-009&\ \ 2.93\ \  &2.361e-011&\ \ 3.86\ \  \\
$968 $ &4.781e-008&\  2.01\ \  &8.546e-010&\ \ 3.10\ \  &1.596e-011&\ \ 4.11\ \  \\
$1152$ &4.036e-008&\  1.95\ \  &6.664e-010&\ \ 2.86\ \  &1.131e-011&\ \ 3.96\ \  \\
$1352$ &3.427e-008&\  2.04\ \  &5.198e-010&\ \ 3.10\ \  &8.228e-012&\ \ 3.97\ \  \\
$1568$ &2.960e-008&\  1.98\ \  &4.125e-010&\ \ 3.12\ \  &6.094e-012&\ \ 4.05\ \  \\
$1800$ &2.575e-008&\  2.02\ \  &3.421e-010&\ \ 2.71\ \  &4.648e-012&\ \ 3.93\ \  \\\hline
  \end{tabular}
  \caption{Numerical errors and convergence orders, for $\parallel\bm u(T)-\bm u_h^T\parallel_{L^2(\Omega)}$ with $\Delta t=10^{-4}$ and $0005$, $Re=10^8$, i.e., $\nu=1.0e-008$; $N$ is the degree of polynomial.}
  \end{table}
\begin{table}[ht]
  \centering
  \begin{tabular}{|c|c|c|c|c|c|c|c|c|}
    \hline
    % after \\: \hline or \cline{col1-col2} \cline{col3-col4}
$K$   &$N=1,error$&$order$      &$N=2,error$&$order$     &$N=3,error$  &$order$\\\hline
$128 $ &1.432e-005 &\      \ \  &5.847e-007&\ \     \ \  & 3.607e-008 &\ \     \ \  \\
$200 $ &8.103e-006 &\  2.55\ \  &2.565e-007&\ \ 3.69\ \  & 1.237e-008 &\ \ 4.80\ \  \\
$288 $ &4.624e-006 &\  3.08\ \  &1.247e-007&\ \ 3.96\ \  & 4.811e-009 &\ \ 5.18\ \  \\
$392 $ &2.877e-006 &\  3.08\ \  &6.510e-008&\ \ 4.22\ \  & 2.424e-009 &\ \ 4.45\ \  \\
$512 $ &1.948e-006 &\  2.92\ \  &3.652e-008&\ \ 4.33\ \  & 1.804e-009 &\ \ 2.21\ \  \\
$648 $ &1.418e-006 &\  2.70\ \  &2.341e-008&\ \ 3.78\ \  & 7.293e-010 &\ \ 7.69\ \  \\
$800 $ &1.021e-006 &\  3.12\ \  &1.552e-008&\ \ 3.90\ \  & 4.626e-010 &\ \ 4.32\ \  \\
$968 $ &7.840e-007 &\  2.77\ \  &1.069e-008&\ \ 3.90\ \  & 2.885e-010 &\ \ 4.95\ \  \\
$1152$ &6.170e-007 &\  2.75\ \  &7.581e-009&\ \ 3.96\ \  & 2.004e-010 &\ \ 4.19\ \  \\
$1352$ &4.868e-007 &\  2.96\ \  &5.509e-009&\ \ 3.99\ \  & 1.402e-010 &\ \ 4.46\ \  \\
$1568$ &4.110e-007 &\  2.28\ \  &4.155e-009&\ \ 3.81\ \  & 1.068e-010 &\ \ 3.67\ \  \\
$1800$ &3.184e-007 &\  3.70\ \  &3.133e-009&\ \ 4.09\ \  & 7.613e-011 &\ \ 4.91\ \  \\ \hline
  \end{tabular}
  \caption{Numerical errors and convergence orders, for $\parallel p(T)-p_h^T\parallel_{L^2(\Omega)}$ with $\Delta t=10^{-4}$ and $0005$, $Re=100$, i.e., $\nu=0.01$; $N$ is the degree of polynomial.}
  \end{table}
  \begin{table}[ht]
  \centering
  \begin{tabular}{|c|c|c|c|c|c|c|c|c|}
    \hline
    % after \\: \hline or \cline{col1-col2} \cline{col3-col4}
$K$     &$N=1,error$&$order$    &$N=2,error$&$order$     &$N=3,error$&$order$\\\hline
$128 $  &1.432e-005&\      \ \  &5.837e-007&\ \     \ \  &3.559e-008 &\ \     \ \  \\
$200 $  &8.103e-006&\  2.55\ \  &2.546e-007&\ \ 3.72\ \  &1.211e-008 &\ \ 4.83\ \  \\
$288 $  &4.624e-006&\  3.08\ \  &1.236e-007&\ \ 3.96\ \  &4.674e-009 &\ \ 5.22\ \  \\
$392 $  &2.844e-006&\  3.16\ \  &6.434e-008&\ \ 4.24\ \  &2.322e-009 &\ \ 4.54\ \  \\
$512 $  &1.921e-006&\  2.94\ \  &3.640e-008&\ \ 4.27\ \  &1.725e-009 &\ \ 2.23\ \  \\
$648 $  &1.388e-006&\  2.78\ \  &2.299e-008&\ \ 3.90\ \  &6.815e-010 &\ \ 7.88\ \  \\
$800 $  &9.945e-007&\  3.17\ \  &1.518e-008&\ \ 3.94\ \  &4.253e-010 &\ \ 4.48\ \  \\
$968 $  &7.591e-007&\  2.83\ \  &1.039e-008&\ \ 3.98\ \  &2.625e-010 &\ \ 5.06\ \  \\
$1152$  &5.926e-007&\  2.85\ \  &7.298e-009&\ \ 4.06\ \  &1.796e-010 &\ \ 4.36\ \  \\
$1352$  &4.645e-007&\  3.04\ \  &5.257e-009&\ \ 4.10\ \  &1.224e-010 &\ \ 4.79\ \  \\
$1568$  &3.895e-007&\  2.38\ \  &3.933e-009&\ \ 3.92\ \  &9.496e-011 &\ \ 3.43\ \  \\
$1800$  &2.982e-007&\  3.88\ \  &2.942e-009&\ \ 4.21\ \  &6.555e-011 &\ \ 5.37\ \  \\\hline
  \end{tabular}
  \caption{Numerical errors and convergence orders, for $\parallel p(T)-p_h^T\parallel_{L^2(\Omega)}$ with $\Delta t=10^{-4}$ and $0005$, $Re=1000$, i.e., $\nu=0.001$; $N$ is the degree of polynomial.}
  \end{table}
 \begin{table}[ht]
  \centering
  \begin{tabular}{|c|c|c|c|c|c|c|c|c|}
    \hline
    % after \\: \hline or \cline{col1-col2} \cline{col3-col4}
 $K$     &$N=1,error$&$order$      &$N=2,error$ &$order$      &$N=3,error$ &$order$\\\hline
$128 $   &1.429e-005 &\ \     \ \  &5.836e-007  &\ \ 3.84\ \  &3.554e-008  &\ \ 5.11\ \ \\
$200 $   &8.057e-006 &\ \ 2.57\ \  &2.544e-007  &\ \ 3.72\ \  &1.208e-008  &\ \ 4.84\ \ \\
$288 $   &4.584e-006 &\ \ 3.09\ \  &1.235e-007  &\ \ 3.96\ \  &4.660e-009  &\ \ 5.22\ \ \\
$392 $   &2.840e-006 &\ \ 3.11\ \  &6.426e-008  &\ \ 4.24\ \  &2.312e-009  &\ \ 4.55\ \ \\
$512 $   &1.918e-006 &\ \ 2.94\ \  &3.638e-008  &\ \ 4.26\ \  &1.716e-009  &\ \ 2.23\ \ \\
$648 $   &1.385e-006 &\ \ 2.76\ \  &2.295e-008  &\ \ 3.91\ \  &6.768e-010  &\ \ 7.90\ \ \\
$800 $   &9.916e-007 &\ \ 3.17\ \  &1.515e-008  &\ \ 3.94\ \  &4.215e-010  &\ \ 4.49\ \ \\
$968 $   &7.564e-007 &\ \ 2.84\ \  &1.036e-008  &\ \ 3.99\ \  &2.602e-010  &\ \ 5.06\ \ \\
$1152$   &5.900e-007 &\ \ 2.86\ \  &7.271e-009  &\ \ 4.07\ \  &1.777e-010  &\ \ 4.38\ \ \\
$1352$   &4.621e-007 &\ \ 3.05\ \  &5.234e-009  &\ \ 4.11\ \  &1.209e-010  &\ \ 4.81\ \ \\
$1568$   &3.872e-007 &\ \ 2.39\ \  &3.912e-009  &\ \ 3.93\ \  &9.402e-011  &\ \ 3.39\ \ \\
$1800$   &2.961e-007 &\ \ 3.89\ \  &2.925e-009  &\ \ 4.21\ \  &6.471e-011  &\ \ 5.41\ \  \\\hline
  \end{tabular}
  \caption{Numerical errors and convergence orders, for $\parallel p(T)-p_h^T\parallel_{L^2(\Omega)}$ with $\Delta t=10^{-4}$ and $0005$, $Re=50000$, i.e., $\nu=0.00002$; $N$ is the degree of polynomial.}
  \end{table}
\begin{table}[ht]
  \centering
  \begin{tabular}{|c|c|c|c|c|c|c|c|c|}  \hline
    % after \\: \hline or \cline{col1-col2} \cline{col3-col4}
$K$      &$N=1,error$&$order$     &$N=2,error$&$order$      &$N=3,error$&$order$ \\\hline
$128 $   &1.429e-005&\ \    \ \   &5.836e-007&\ \    \ \   &3.554e-008&\ \     \ \  \\
$200 $   &8.057e-006&\ \ 2.57\ \  &2.544e-007&\ \ 3.72\ \  &1.208e-008&\ \ 4.84\ \  \\
$288 $   &4.584e-006&\ \ 3.09\ \  &1.235e-007&\ \ 3.96\ \  &4.659e-009&\ \ 5.22\ \  \\
$392 $   &2.840e-006&\ \ 3.11\ \  &6.426e-008&\ \ 4.24\ \  &2.311e-009&\ \ 4.55\ \  \\
$512 $   &1.918e-006&\ \ 2.94\ \  &3.638e-008&\ \ 4.26\ \  &1.716e-009&\ \ 2.23\ \  \\
$648 $   &1.385e-006&\ \ 2.76\ \  &2.295e-008&\ \ 3.91\ \  &6.767e-010&\ \ 7.90\ \  \\
$800 $   &9.915e-007&\ \ 3.17\ \  &1.514e-008&\ \ 3.95\ \  &4.215e-010&\ \ 4.49\ \  \\
$968 $   &7.564e-007&\ \ 2.84\ \  &1.036e-008&\ \ 3.98\ \  &2.601e-010&\ \ 5.07\ \  \\
$1152$   &5.899e-007&\ \ 2.86\ \  &7.270e-009&\ \ 4.07\ \  &1.776e-010&\ \ 4.38\ \  \\
$1352$   &4.621e-007&\ \ 3.05\ \  &5.233e-009&\ \ 4.11\ \  &1.209e-010&\ \ 4.80\ \  \\
$1568$   &3.871e-007&\ \ 2.39\ \  &3.912e-009&\ \ 3.93\ \  &9.400e-011&\ \ 3.40\ \  \\
$1800$   &2.960e-007&\ \ 3.89\ \  &2.924e-009&\ \ 4.22\ \  &6.470e-011&\ \ 5.41\ \  \\ \hline
  \end{tabular}
  \caption{Numerical errors and convergence orders, for $\parallel p(T)-p_h^T\parallel_{L^2(\Omega)}$ with $\Delta t=10^{-4}$ and $0005$, $Re=1.0e+008$, i.e., $\nu=1.0e-008$; $N$ is the degree of polynomial.}
  \end{table}

In Figures 1-4, we numerically show one of the striking benefits of the proposed schemes: the conditional number of the matrix of the corresponding matrix equation does not increase with the refining of the meshes for a wide range of Reynolds numbers. The well symmetric properties of the schemes should make contribution to this benefit. The $N$ denotes the degree of polynomial.

%In our experiment we discovered that the conditional number of the left-hand side matrix is different. The conditional number became smaller when step h became smaller (the reciprocal of h became bigger as in figures ) until stayed in a line level. In many Discontinuous Galerkin Methods the conditional number in solving other equations will become bigger when the step h decreases. From this point, our scheme is better.  As usual, the conditional number of left side stiffness matrix became big until we can not simulate the equations along with the step h refined. But in our experiments we found the different situation. We represent the figures of the relations of reciprocal of step h and conditional numbers to display the beautiful results. Perhaps we can guess that the conditional number of stiffness matrix in Discontinuous Methods is related not only the refined step h but also the schemes used.
%, we fixed the approximate polynomial degree N=1,2,3, and give changes of the conditional number of
%stiffness matrix with the different Reynold numbers. We can see the difference in them.
%%%%%%%%%%%%%%%%%%%%%%%%%%%%%%%%%%%%%%%%%%%%%%%%%
\begin{figure}[ht]
\begin{minipage}[t]{0.49\linewidth}
\centering
\includegraphics[width=2.3in,height=2.6in]{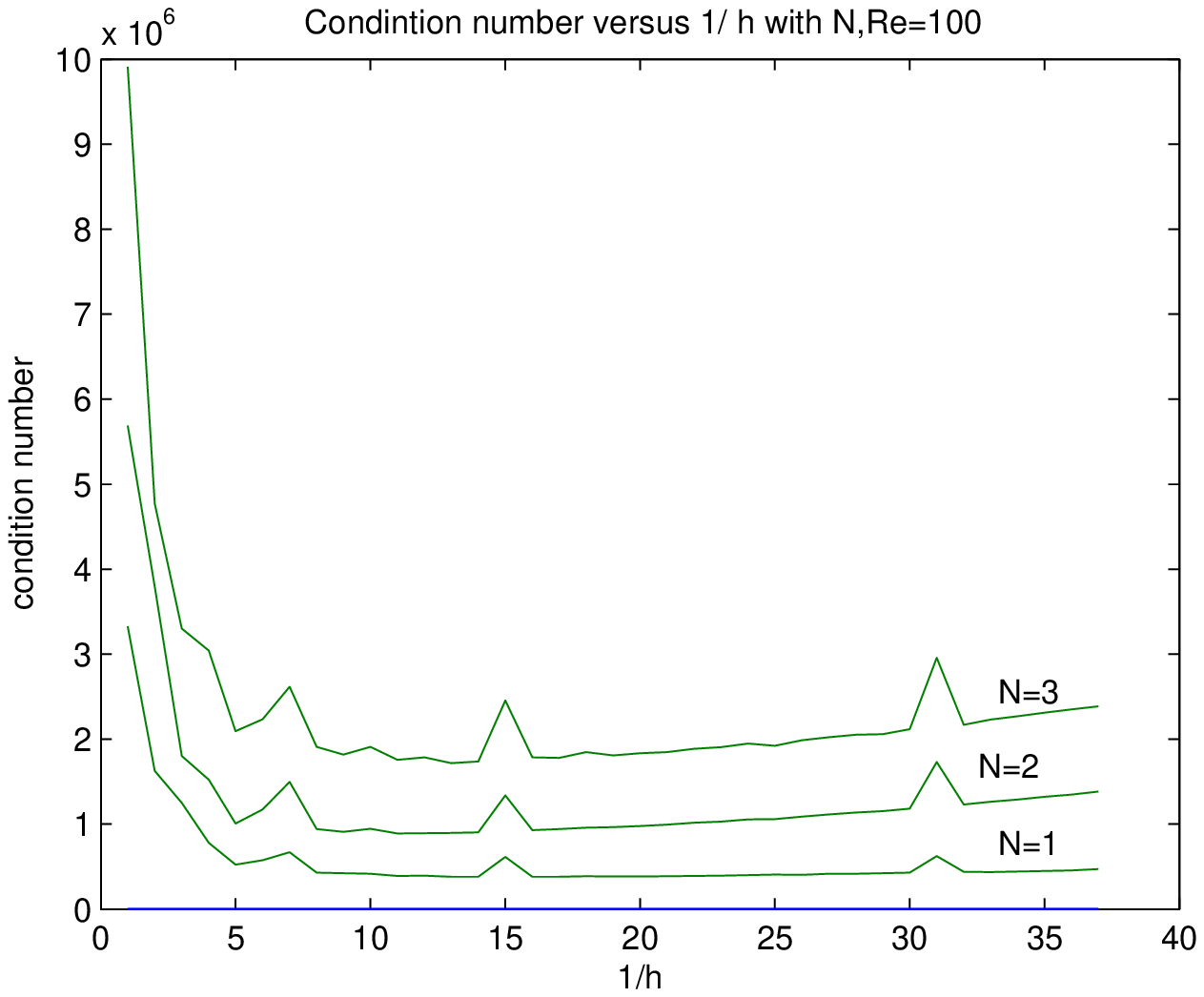}
\caption{Conditional numbers vs $1/h$.}
\end{minipage}
\begin{minipage}[t]{0.49\linewidth}
\centering
\includegraphics[width=2.3in,height=2.6in]{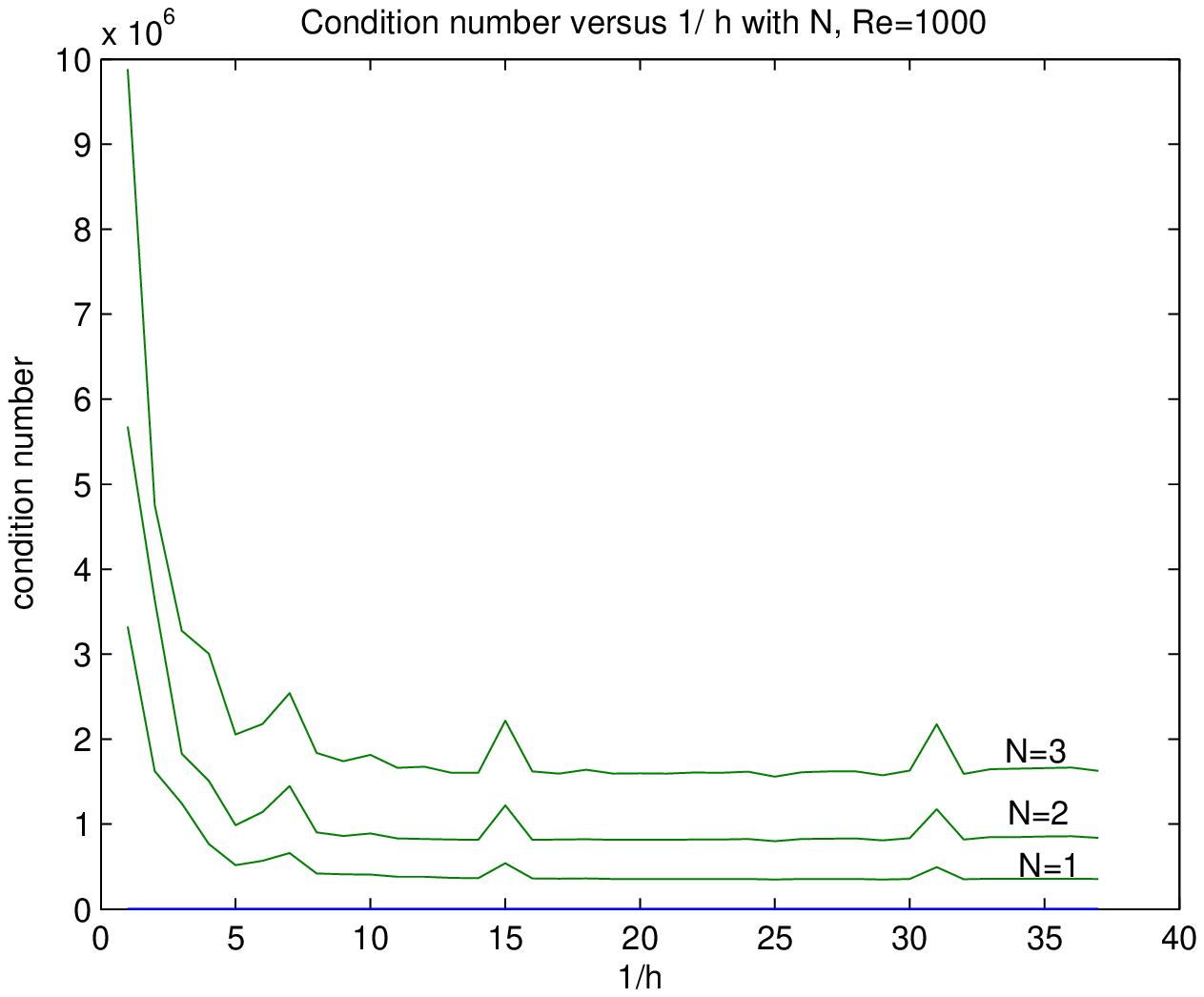}
\caption{Conditional numbers vs $1/h$.}
\end{minipage}
\end{figure}
%%%%%%%%%%%%%%%%%%%%%%%%%%%%%%%%%%%%%%%%%%%%%%%%%%%%%%
\begin{figure}[ht]
\begin{minipage}[t]{0.49\linewidth}
\centering
\includegraphics[width=2.3in,height=2.6in]{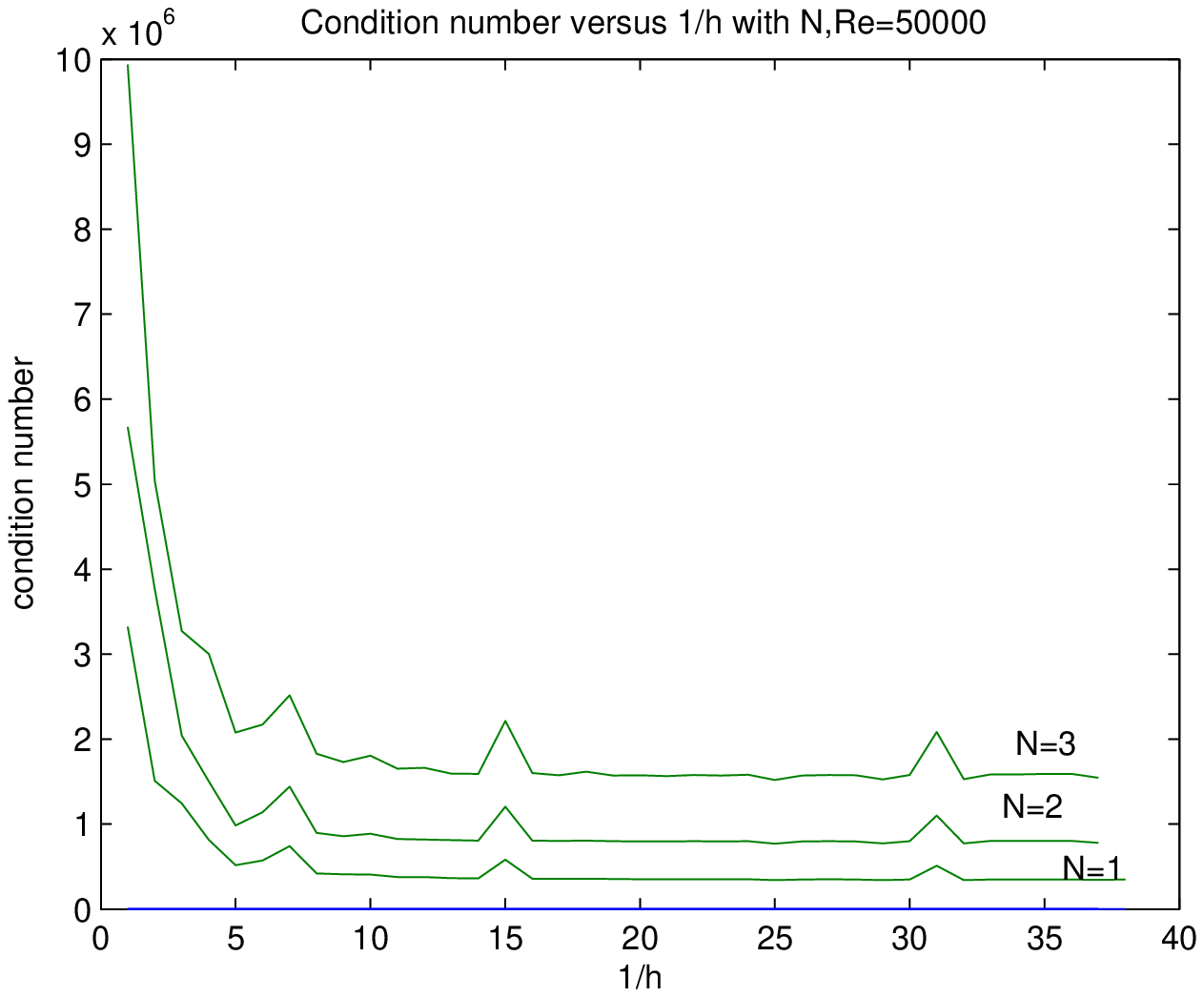}
\caption{Conditional numbers vs $1/h$.}
%\label{fig:side:a}
\end{minipage}
\begin{minipage}[t]{0.49\linewidth}
\centering
\includegraphics[width=2.3in,height=2.6in]{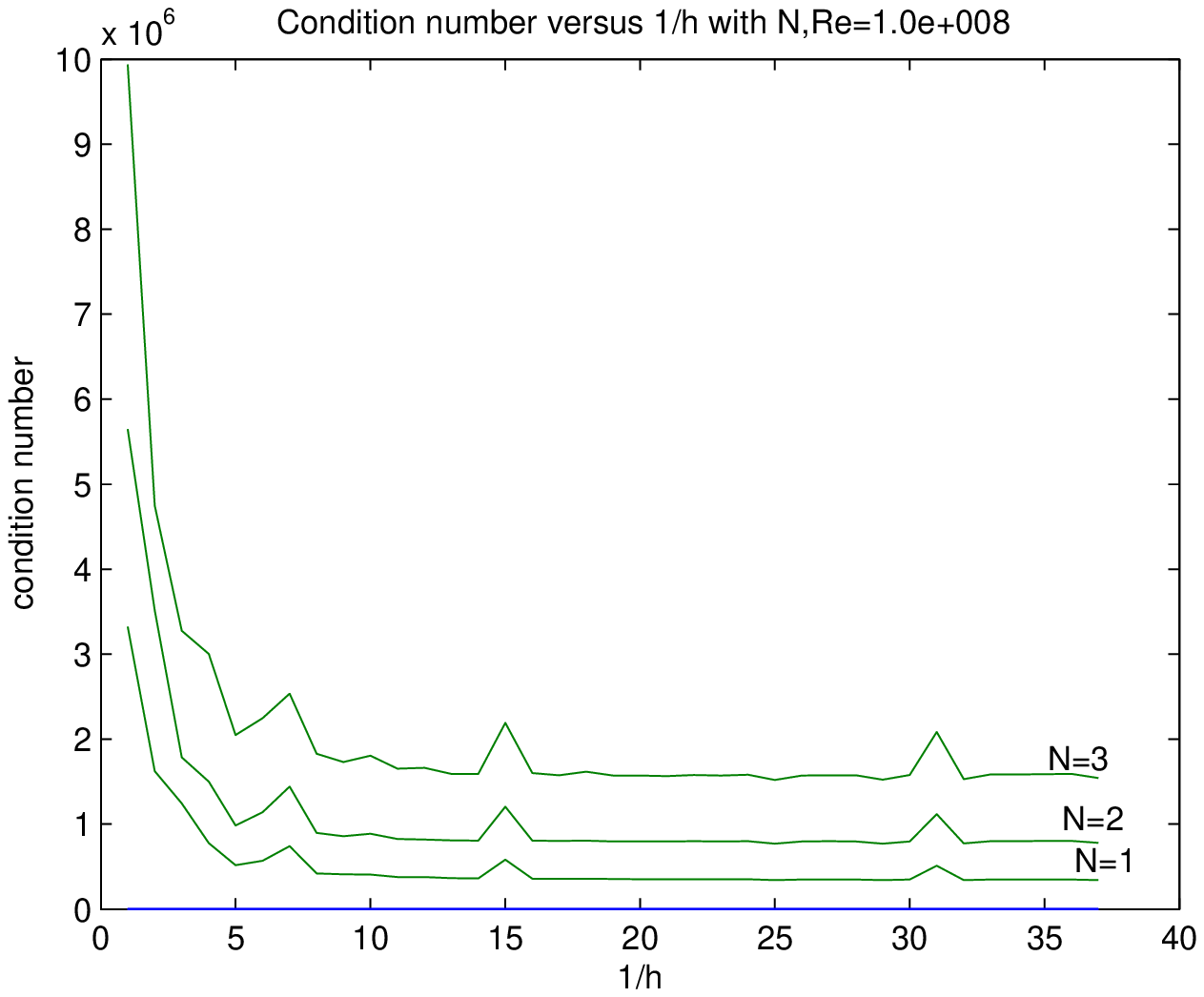}
\caption{Conditional numbers vs $1/h$.}
%\label{fig:side:b}
\end{minipage}
\end{figure}

\vskip 0.2cm
\noindent\textbf{Example 5.2.} For the convenience, we still choose the domain $\Omega=[0,1]\times[0,1]$; the source term $\bm f=(f_1,f_2)$, and the initial and boundary conditions are given as follows:
\begin{equation}
\left\{ \begin{array}
 {l@{\quad} l}
 f_1(x,y,t)=e^{-\nu(x+y)t},\\
 f_2(x,y,t)=sin(\nu(x+y)t);
 \end{array}
 \right.
\end{equation}
\begin{align}
\bm u(x,y,t)=\bm 0,\ \forall (x,y,t)\in\partial\Omega\times[0,T]; \ \bm u(x,y,0)=\bm 0,\ \forall (x,y)\in\Omega.
\end{align}
In this simulation, we use the first order polynomial, the spatial stepsize $h=1/8$, and the Reynolds number $Re=100$.
Figures 5-13 show the contours of ${u_1}_h$, ${u_2}_h$, and $p_h(t)$ for the different times $t=0.001,\,0.01,\,0.1$.

%We choose different time (i.e. $t=0.001,0.01,0.1$) to simulate the exact velocity and the pressure $(\bm u(t),p(t))$.
%Obviously, the numerical solutions changed with different times. In Figures 5,6,7,8,9,10,11,12,13, we show the contour figures
%about the exact solutions with changed time.
%%%%%%%%%%%%%%%%%%%%%%%%%%%%%%%%%%%%%%%%%%%%%%%%%
\begin{figure}[ht]
\begin{minipage}[t]{0.49\linewidth}
\centering
\includegraphics[width=2.3in,height=2.3in]{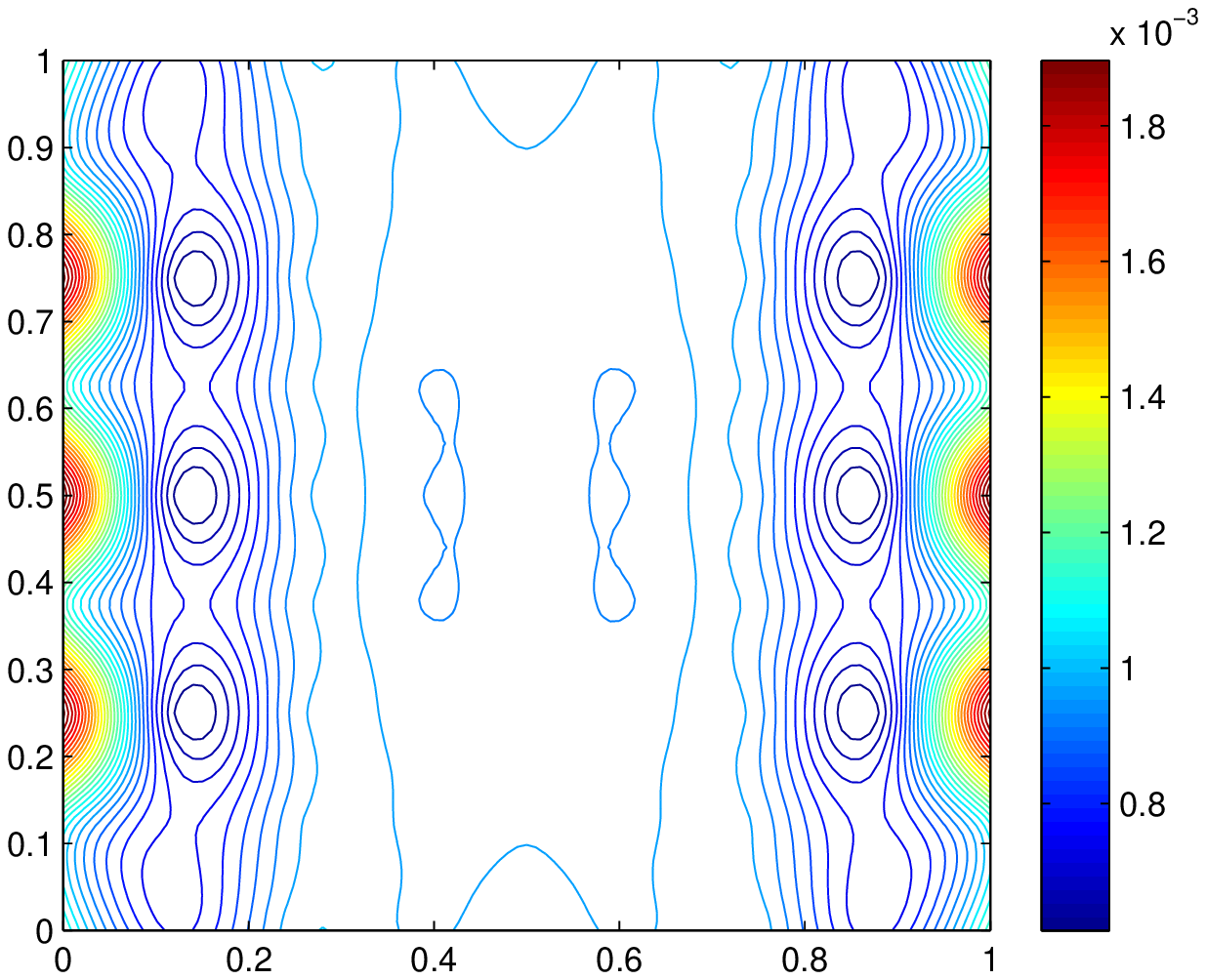}
\caption{The contour of ${u_1}_h$ with 001.}
%\label{}
\end{minipage}
\begin{minipage}[t]{0.49\linewidth}
\centering
\includegraphics[width=2.3in,height=2.3in]{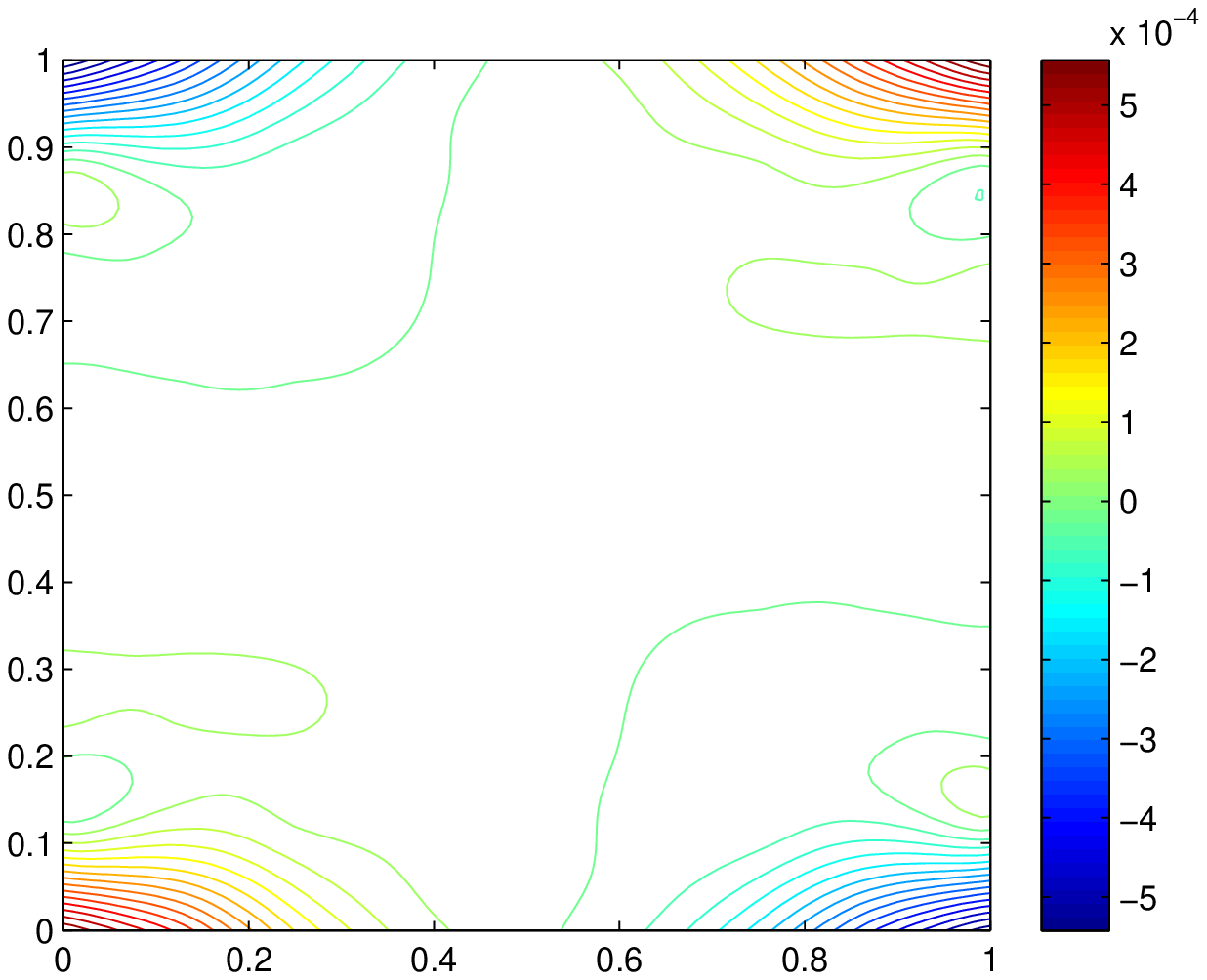}
\caption{The contour of ${u_2}_h$ with t=0.001.}
%\label{}
\end{minipage}
\end{figure}
%%%%%%%%%%%%%%%%%%%%%%%%%%%%%%%%%%%%%%%%%%%%%%%%%%%%%%
\begin{figure}[ht]
\begin{minipage}[t]{0.49\linewidth}
\centering
\includegraphics[width=2.3in,height=2.3in]{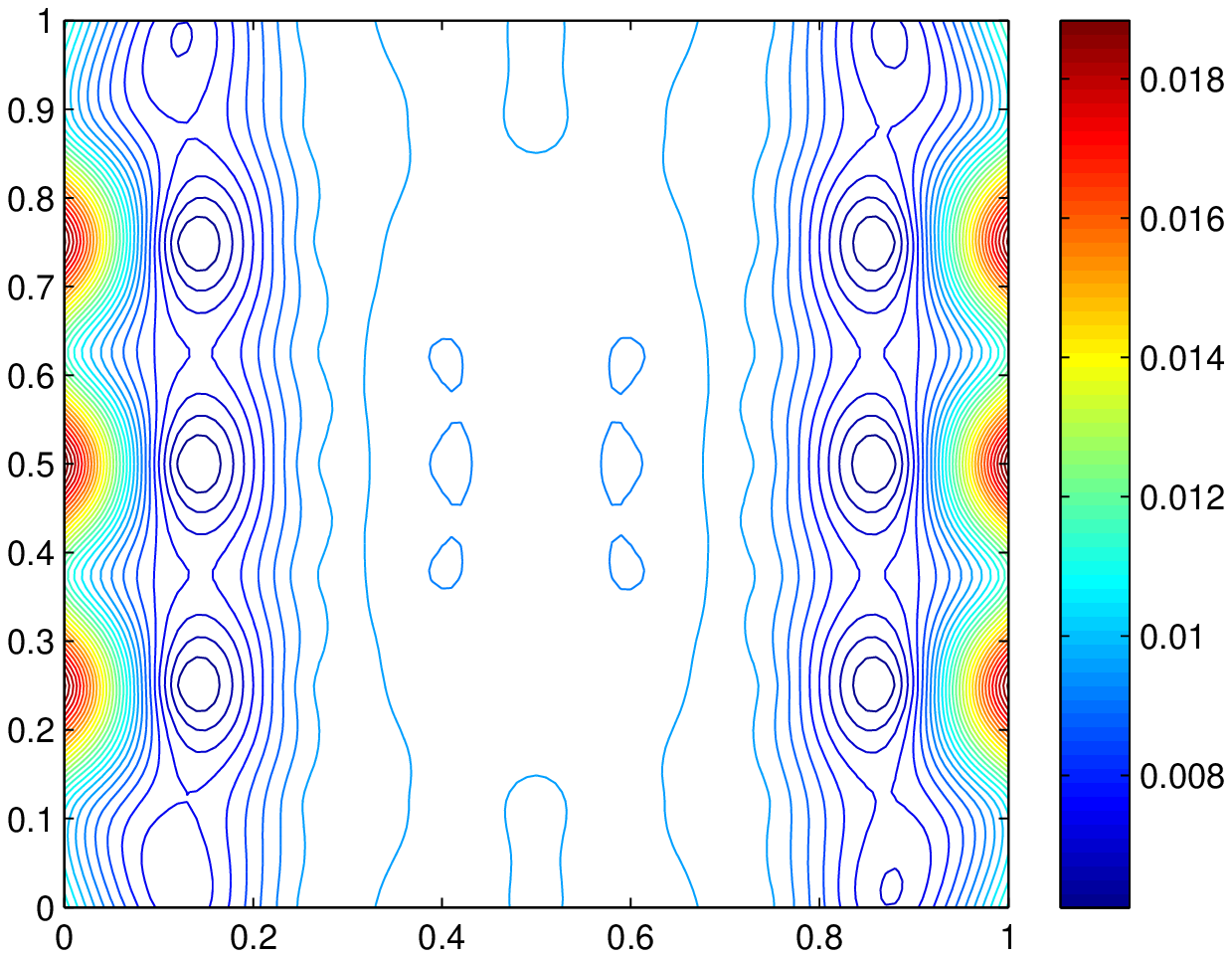}
\caption{The contour of ${u_1}_h$ with t=0.01.}
%\label{fig:side:a}
\end{minipage}
\begin{minipage}[t]{0.49\linewidth}
\centering
\includegraphics[width=2.3in,height=2.3in]{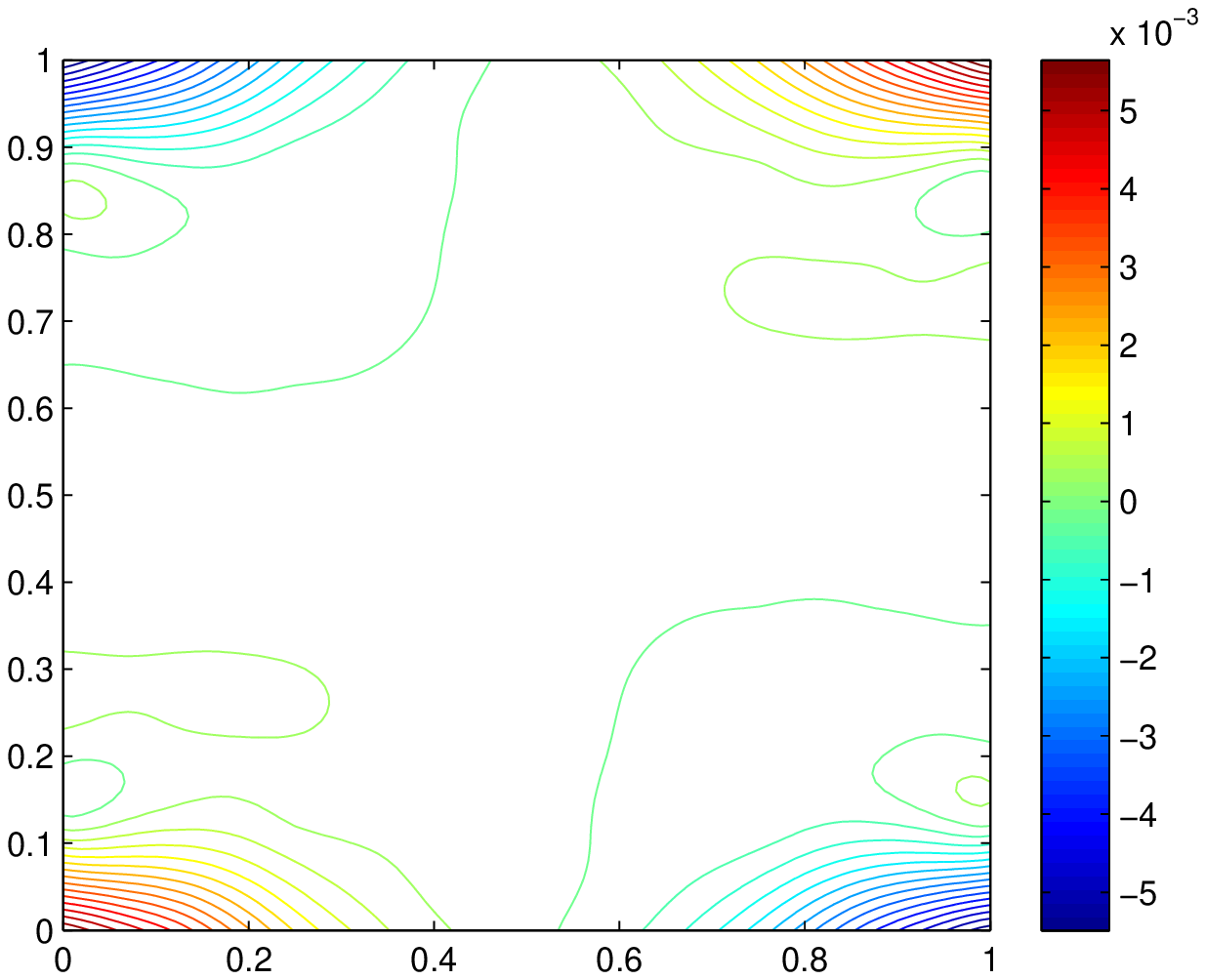}
\caption{The contour of ${u_2}_h$ with t=0.01.}
%\label{fig:side:b}
\end{minipage}
\end{figure}
%%%%%%%%%%%%%%%%%%%%%%%%%%%%%%%%%%%%%%%%%%%%%%%%%%%%%%
\begin{figure}[ht]
\begin{minipage}[t]{0.49\linewidth}
\centering
\includegraphics[width=2.3in,height=2.3in]{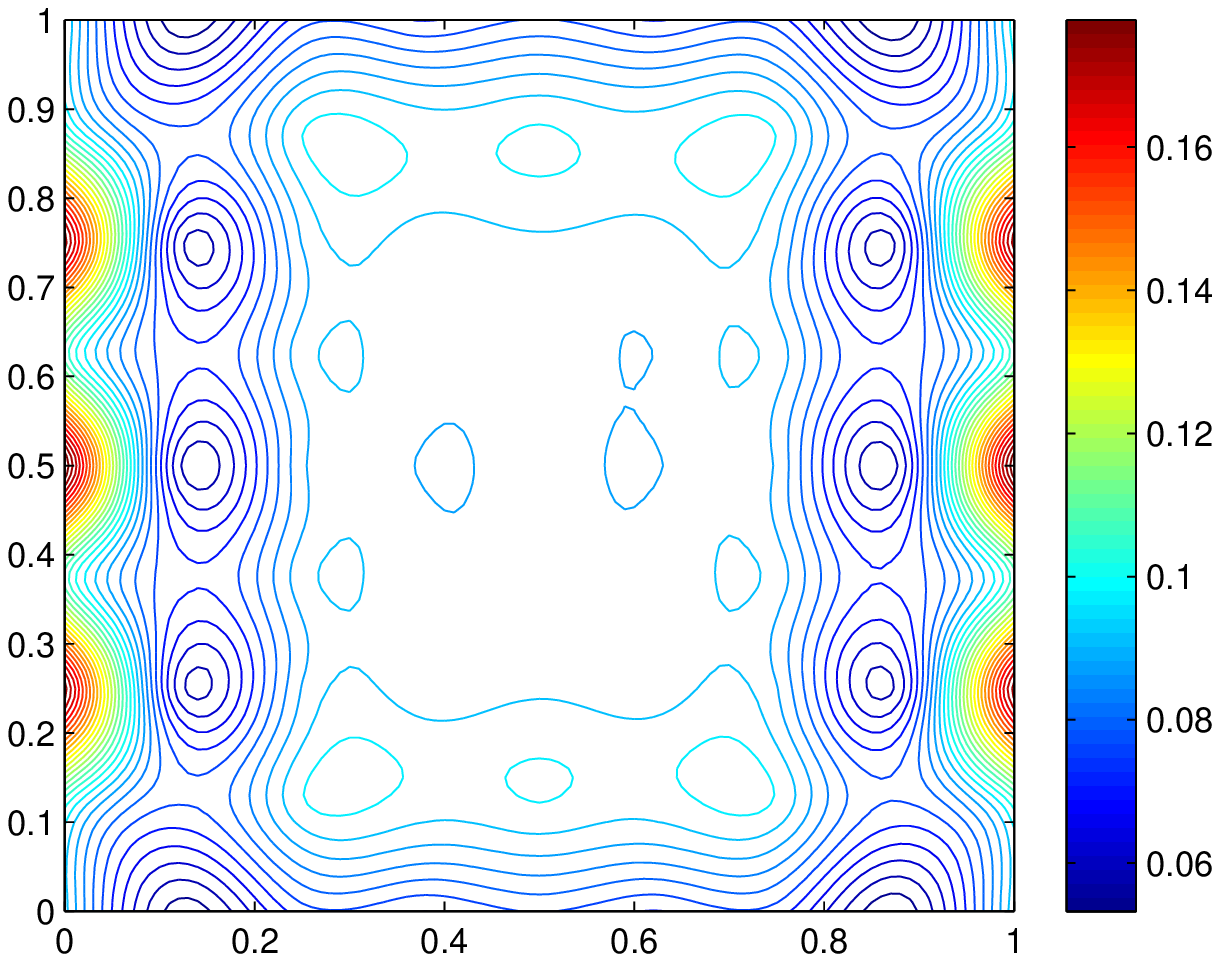}
\caption{The contour of ${u_1}_h$ with t=0.1.}
%\label{fig:side:a}
\end{minipage}
\begin{minipage}[t]{0.44\linewidth}
\centering
\includegraphics[width=2.3in,height=2.3in]{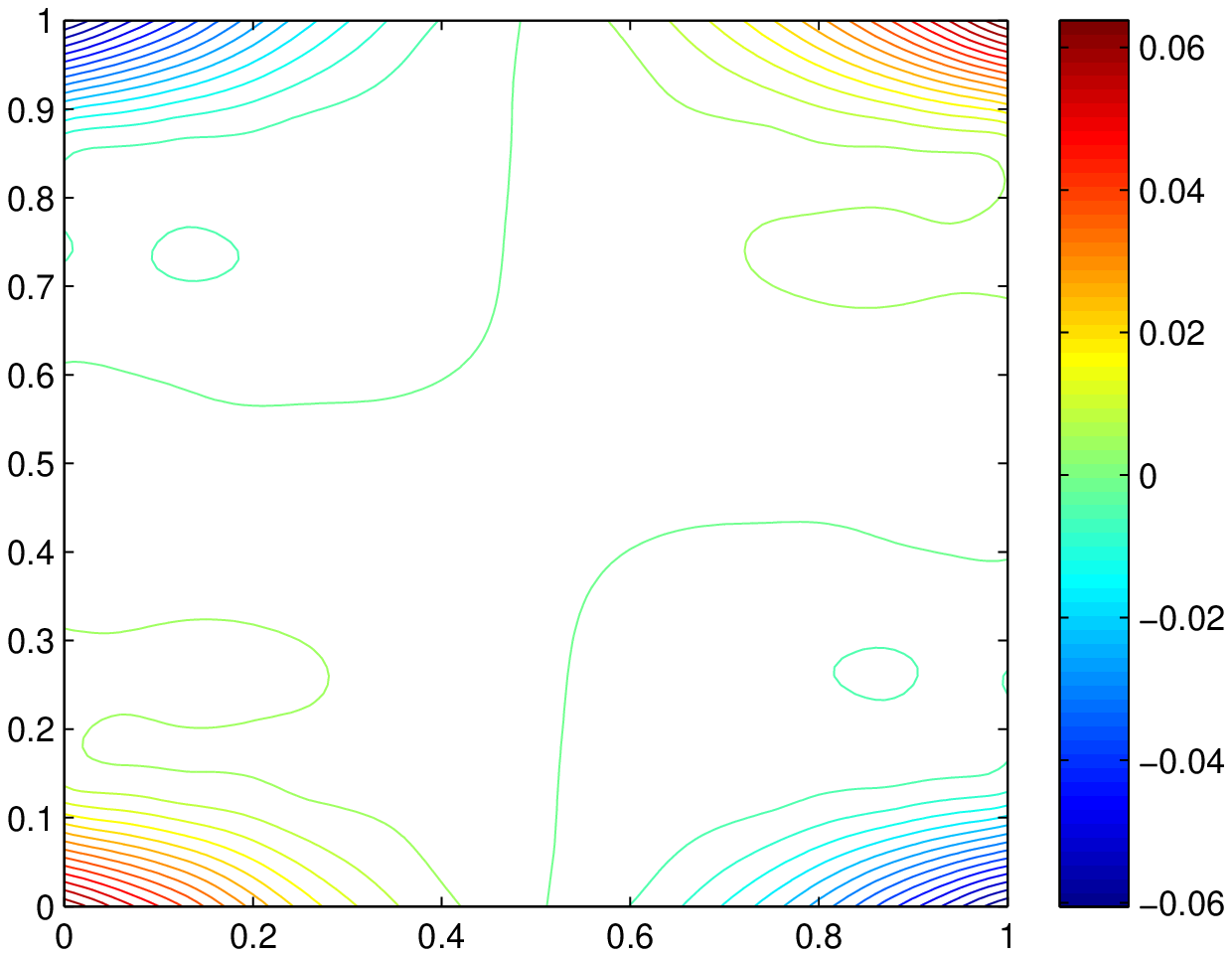}
\caption{The contour of ${u_2}_h$ with t=0.1.}
%\label{fig:side:b}
\end{minipage}
\end{figure}
%%%%%%%%%%%%%%%%%%%%%%%%%%%%%%%%%%%%%%%%%%%%%%%%%%%%%
\begin{figure}[ht]
\begin{minipage}[t]{0.49\linewidth}
\centering
\includegraphics[width=2.4in,height=2.3in]{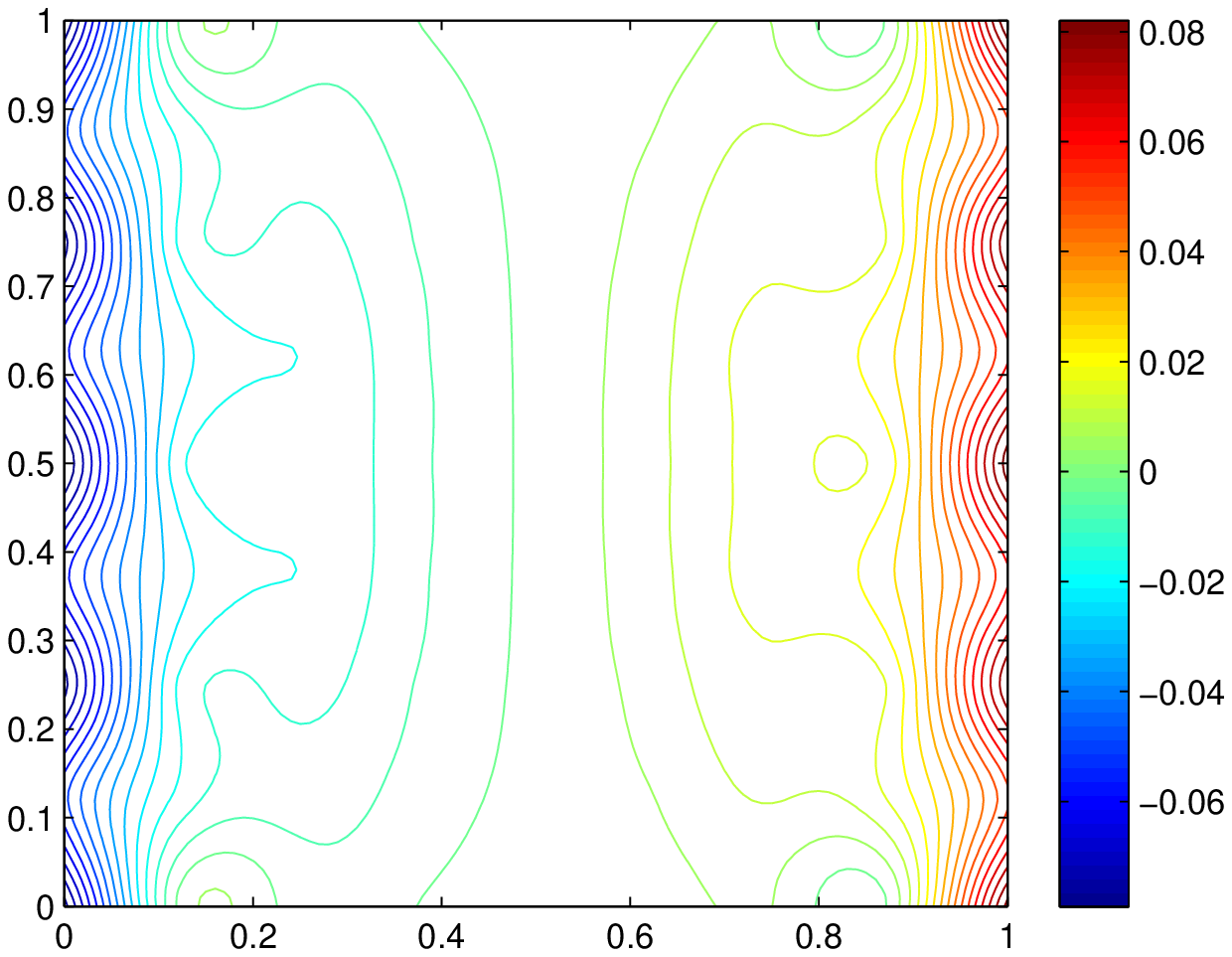}
\caption{The contour of $p_h$ with t=0.001.}
%\label{}
\end{minipage}
\begin{minipage}[t]{0.49\linewidth}
\centering
\includegraphics[width=2.4in,height=2.3in]{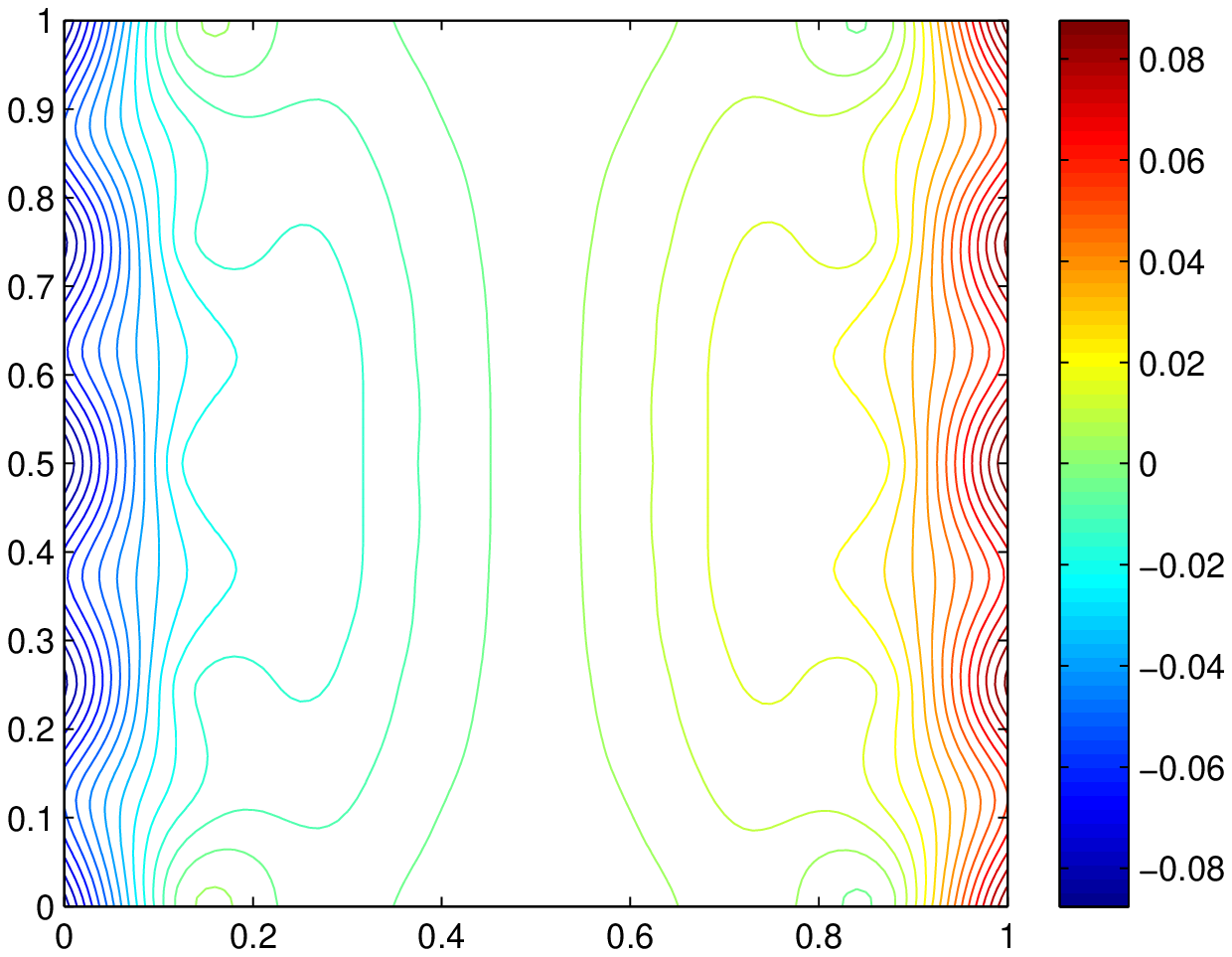}
\caption{The contour of $p_h$ with t=0.01.}
%\label{}
\end{minipage}
\end{figure}
%%%%%%%%%%%%%%%%%%%%%%%%%%%%%%%%%%%%%%%%%%%%%%%%%%%%%%
\begin{figure}[ht]
\centering
\includegraphics[width=2.4in,height=2.3in]{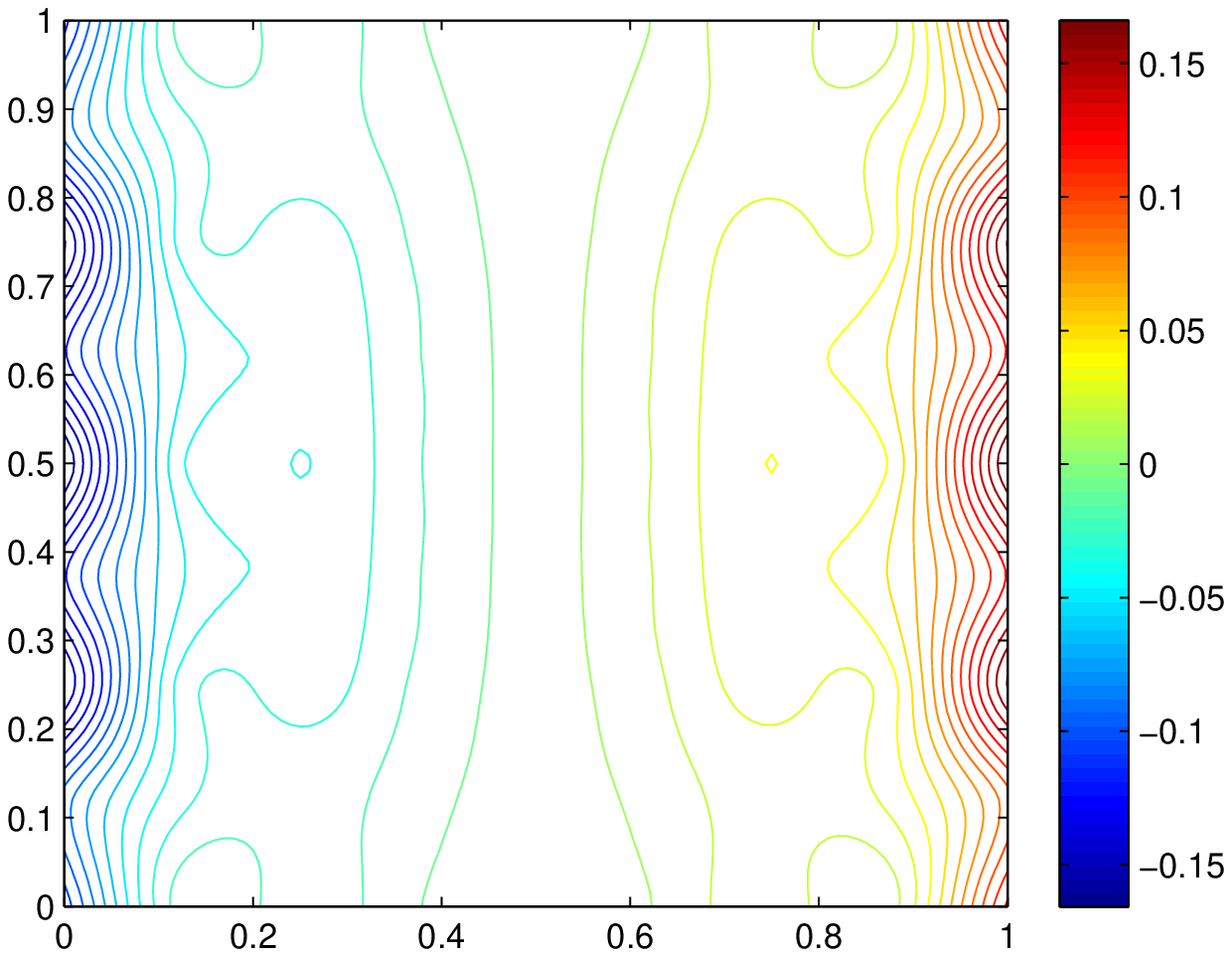}
\caption{The contour of $p_h$ with t=0.1.}
%\label{fig:side:a}
\end{figure}
%%%%%%%%%%%%%%%%%%%%%%%%%%%%%%%%%%%%%%%%%%%%%%%%%%%%%%
\section{Conclusions}
By carefully constructing the numerical fluxes, adding the penalty terms, and using the characteristic method to discretize the time derivative and nonlinear convective term, we design the effective LDG scheme to solve the time-dependent
convection-dominated Navier-Stokes equations in $\mathbb{R}^2$. Besides the general advantages of the LDG scheme, the proposed scheme is theoretically proved or numerically verified to have the following benefits: 1) it is well symmetric, so easy to do theoretical analysis and numerical computation; 2) theoretically proved to unconditionally stable; 3) numerically verified to have the optimal convergence orders; 4) the conditional number of the matrix of the corresponding matrix equation does not increase with the refining of the meshes; 5) the scheme is efficient for a wide range of Reynolds numbers. The other possible variational formulation is presented in the Appendix.

%Combining the characteristic method and the local discontinuous
%Galerkin method with carefully constructing numerical fluxes, we
%design the variational formulations for the time-dependent
%convection-dominated Navier-Stokes equations in $\mathbb{R}^2$. The
%proposed symmetric variational formulation is strictly proved to be
%unconditionally stable; and the scheme has the striking benefit that the conditional number of the matrix of the corresponding matrix equation does not increase with the refining of the meshes. The presented scheme works well for a wide
%range of Reynolds numbers, e.g., the scheme still has good error
%convergence when $Re=0.5 e+005$ or $1.0 e+ 008$. Extensive
%numerical experiments are performed to show the optimal convergence
%orders and the contours of the solutions of the equation with given
%initial and boundary conditions.

\section*{Appendix}
Here we present the other variational formulation for the time-dependent
convection-dominated Navier-Stokes equations. We still use the characteristic method to discretize the time derivative and the nonlinear convection term.
%\subsection*{The second variational formulation}
 Multiplying the first, second, and
third equation of (\ref{2.3p}) by arbitrary smooth test functions $\bm v, \bar{\bm \tau}, q$, respectively, and integrating by parts over an arbitrary subset $E\in\mathscr{E}_h$ twice, we obtain
\begin{equation}
\left\{ \begin{array}
 {l@{\quad} l}
 \int_E(\bm u_t+(\bm u\cdot\nabla)\bm u)\bm v-\int_E\sqrt{\nu}\nabla\cdot\bar{\bm\sigma}\bm v +\int_{\partial
 E}\sqrt{\nu}(\bar{\bm\sigma}-\bar{\bm\sigma}^{\ast})\cdot\bm v\cdot\bm n_E\\
 +\int_E \nabla p\cdot\bm v-\int_{\partial E}(p-p^{\ast})\bm v\cdot\bm n_E
=\int_E\bm f\bm v,\ \forall \bm v\in\bm V,\\
\\
\int_E\bar{\bm\sigma}:\bar{\bm\tau}+\int_E\sqrt{\nu}\nabla\cdot\bar{\bm\tau}\cdot\bm u-\int_{\partial E}\sqrt{\nu}\bm u^{\ast}\cdot\bar{\bm\tau}\cdot\bm n_E=0,\ \forall \bar{\bm\tau}\in\bm V^2,\\
\\
 -\int_E\nabla q\cdot\bm u+\int_{\partial E} q\bm u^{\ast\ast}\cdot\bm n_E =0,\ \forall q\in M,
 \end{array}
 \right.
\end{equation}
where $\bm n_E$ is the outward unit normal to $\partial E$. Choosing the fluxes of $\bar{\bm\sigma}^{\ast},\bm u^{\ast},\bm u^{\ast\ast},p^{\ast}$ as:
$$ \bar{\bm\sigma}^{\ast}=\{\{\bar{\bm\sigma}\}\},\ \ \bm u^{\ast}=\{\{\bm u\}\},\ \
 p^{\ast}=\{\{p\}\},\ \ \bm u^{\ast\ast}=\{\{\bm u\}\},$$
leads to the second variational formulation:
\begin{equation}
\left\{ \begin{array}
 {l@{\quad} l}
 \sum_{E\in\mathscr{E}_h}(\bm u_t+(\bm u\cdot\nabla)\bm u,\bm v)_E
  -\sum_{E\in\mathscr{E}_h}(\sqrt{\nu}\nabla\cdot\bar{\bm\sigma},\bm v)_E\\
  +\sum_{e\in\mathscr{E}_h^B}(\sqrt{\nu}\lbrack\bar{\bm\sigma}\rbrack,\{\{\bm v\}\}\otimes\bm n_e)_e
  +\sum_{E\in\mathscr{E}_h}(\nabla p,\bm v)_E\\
  -\sum_{e\in\mathscr{E}_h^B}(\lbrack p\rbrack,\{\{\bm v\}\}\cdot\bm n_e)_e=\sum_{E\in\mathscr{E}_h}(\bm f,\bm v)_E,\\
  \\
 %\cr\noalign{\vskip  0.01 mm}
 \sum_{E\in\mathscr{E}_h}(\bar{\bm\sigma},\bar{\bm\tau})_E
 +\sum_{E\in\mathscr{E}_h}(\sqrt{\nu}\nabla\cdot\bar{\bm\tau},\bm u)_E\\-\sum_{e\in\mathscr{E}_h^B}(\sqrt{\nu}\lbrack\bar{\bm\tau}\rbrack,\{\{\bm u\}\}\otimes\bm n_e)_e=0,\\
 \\
-\sum_{E\in\mathscr{E}_h}(\nabla q,\bm u)_E+\sum_{e\in\mathscr{E}_h^B}(\lbrack q\rbrack,\{\{\bm u\}\}\cdot\bm n_e)_e=0.
 \end{array}
 \right.
\end{equation}
This formulation is unconditionally stable as the first one; and it is the strong formulation of the first one.

\noindent

%\section*{Acknowledgment}


\begin{thebibliography}{00}
%
\bibitem{r1}T. Arbogast, M.F. Wheeler, A characteristics-mixed finite element method for advection-dominated transport problems, {\it SIAM J. Numer. Anal.}, {\bf 32(2)} (1995) 404-424.

\bibitem{r2}K. Boukir, Y. Maday, B. M\'{e}tivet, A high order characteristics method for the
 incompressible Navier-Stokes equations, {\it Comput. Methods Appl. Mech. Engrg.}, {\bf 116} (1994) 211-218.

\bibitem{r3}K. Boukir, Y. Maday, B. M\'{e}tivet, A high order characteristics$/$finite element method for the incompressible Navier-Stokes equations, {\it Int. J.  Numer. Methods Fluids}, {\bf 25} (1997) 1421-1454.

\bibitem{r4} B. Cockburn, C.-W. Shu, The local discontinuous Galerkin method for time-dependent convection-diffusion systems, {\it SIAM J. Numer. Anal.}, {\bf 35(6)} (1998) 2440-2463.

\bibitem{r5} B. Cockburn, G. Kanschat, D. Sch\"{o}tzau, C. Schwab, Local discontinuous Galerkin methods for the Stokes System, {\it SIAM J. Numer. Anal.}, {\bf 40(1)} (2003) 319-343.

\bibitem{r6} B. Cockburn, G. Kanschat, D. Sch\"{o}tzau, The local discontinuous Galerkin method for the Ossen equations, {\it Math. Comput.}, {\bf 73} (2004) 569-593.

\bibitem{r7}B. Cockburn, G. Kanschat, D.Sch\"{o}tzau, A locally conservative LDG method for the incompressible Navier-Stokes equations, {\it Math. Comput.}, {\bf 74} (2005) 1067-1095.

\bibitem{r8}B. Cockburn, G. Kanschat, D. Sch\"{o}tzau, A note on discontinuous Galerkin divergence-free solutions of Navier-Stokes equations, {\it J. Sci. Comput.}, {\bf 31} (2007) 61-73.

\bibitem{r9}P. Castillo, B. Cockburn, I. Perugia, D. Sch\"{o}tzau, An apriori error analysis of
the local discontinuous Galerkin method for elliptic problems, {\it SIAM J. Numer. Anal.}, {\bf 38} (2001) 1676-1706.

\bibitem{r10} P. Castillo, B. Cockburn, C. Schwab, Optimal apriori error analysis for the hp-version of the local discontinuous Galerkin method for convection-diffusion problems, {\it Math. Comput.}, {\bf 71} (2002) 455-478.

\bibitem{r11} V. Girautl, R. Scott, A quasi-local interpolation operator preserving the discrete divergence, {\it Calcolo}, {\bf 40} (2003) 1-19.

\bibitem{r12} V. Girautl, B. Rivi\`{e}re, M.F. Wheeler, A splitting method using discontinuous Galerkin for the transient incompressible Navier-Stokes equations,  {\it ESAIM: M2AN}, {\bf 39} (2005) 1115-1147.

\bibitem{r13}V. Girautl, P. Raviart, {\it Finite Element Methods for Navier-Stokes Equations: Theory and Algorithms}, Volume 5 of Springer Series in Computational Mathematics, (Springer-Verlag, Berlin, 1986).

\bibitem{r14}B. Rivi\`{e}re, V. Girault, Discontinuous finite element methods for incompressible flows on subdomains with non-matching interfaces, {\it Comput. Meth. Appl. Mech. Engrg.}, {\bf 195} (2006) 3274-3292.

\bibitem{r15}B. Rivi\`{e}re, {\it Discontinuous Galerkin Methods for Solving  Elliptic and Parabolic Equations: Theory and Implementation}, Society for Industrial and Applied Mathematics, (SIAM, 1999).

\bibitem{r16} R. Temam, {\it Navier-Stokes Equations: Theory and Numerical analysis}, North-Holland-Amsterdam. New York. Oxford, (Elsevier Science Publishers B.V., 1984).

\bibitem{r17} F. Bassi and S. Rebay, A high-order accurate discontinuous finite elemnet method for the numerical solution of the compressible Navier-Stokes equations, {\it J. Comput. Phys.}, {\bf 131} (1997) 267-279.

%\bibitem{r18} V.J. Ervin, N. Heuer, J.P. Roop, Numerical approximation of a time dependent, nonlinear, space-fractional diffusion equation, {\it SIAM J. Numer. Anal.}, {\bf 45(2)} (2007) 572-591.

\bibitem{r19} W.H. Deng, Finite element method for the space and time fractional Fokker-Plank equations, {\it SIAM J. Numer. Anal}, {\bf 47(1)} (2008) 204-226.

\bibitem{r20} W.H. Deng, J.S. Hesthaven,  Local discontinuous Galerkin methods for fractional diffusion equations, {\it ESAIM: M2AN}, {\bf 47}
(2013) 1845-1864.

\bibitem{r21} R.H. Nochetto, J.-H. Pyo, The Gauge-Uzawa finite element method. Part I: the Navier-Stokes equations, {\it SIAM J. Numer. Anal.}, {\bf  43(3)} (2005) 1043-1068 .

\bibitem{r22} Z.X. Chen, Characteristic mixed discontinuous finite element methods for advection-dominated diffusion problems, {\it Comput. Methods Appl. Mech. Engrg.}, {\bf 191} (2002) 2509-2538.

\bibitem{r23} P.G. Ciarlet, {\it The Finite Element Method for Elliptic Problems},  North-Holland-Amsterdam. New York. Oxford, (Elsevier Science Publishers B.V., 1978).

\bibitem{r24} Y. He, A fully discrete stabilized finite-element method for the time-dependent Navier-Stokes problem, {\it IMA Journal of Numerical Analysis} {\bf 23} (2003) 665-691.
\end{thebibliography}
\end{document}